\documentclass[10pt,journal]{IEEEtran}

\usepackage{nomencl}
\makenomenclature

\renewcommand{\IEEEQED}{\IEEEQEDclosed}
\usepackage{color}

\usepackage{subfigure}
\usepackage{graphicx,cite,epsfig,amssymb,amsmath,multirow,lettrine,flushend,extarrows}
\usepackage{algorithm}
\usepackage{algorithmic}

\begin{document}
%
\title{Online Energy Management for a Sustainable Smart Home with an HVAC Load and Random Occupancy}

\author{{Liang~Yu,~\IEEEmembership{Member,~IEEE}, Tao~Jiang,~\IEEEmembership{Senior Member,~IEEE}, and Yulong Zou,~\IEEEmembership{Senior Member,~IEEE}}
\thanks{\newline L. Yu and Y. Zou are with Key Laboratory of Broadband Wireless Communication and Sensor Network Technology of Ministry of Education, Nanjing University of Posts and Telecommunications, Nanjing 210003, P. R. China. (email: \{liang.yu,yulong.zou\}@njupt.edu.cn) \newline
T. Jiang is Wuhan National Laboratory for Optoelectronics, School of Electronics Information and Communications, Huazhong University of Science and Technology, Wuhan 430074, P. R. China. (email: Tao.Jiang@ieee.org) \newline
}}


\maketitle

\begin{abstract}
In this paper, we investigate the problem of minimizing the sum of energy cost and thermal discomfort cost in a long-term time horizon for a sustainable smart home with a Heating, Ventilation, and Air Conditioning (HVAC) load. Specifically, we first formulate a stochastic program to minimize the time average expected total cost with the consideration of uncertainties in electricity price, outdoor temperature, renewable generation output, electrical demand, the most comfortable temperature level, and home occupancy state. Then, we propose an online energy management algorithm based on the framework of Lyapunov optimization techniques without the need to predict any system parameters. The key idea of the proposed algorithm is to construct and stabilize four queues associated with indoor temperature, electric vehicle charging, and energy storage. Moreover, we theoretically analyze the feasibility and performance guarantee of the proposed algorithm. Extensive simulations based on real-world traces show the effectiveness of the proposed algorithm.
\end{abstract}

\IEEEpeerreviewmaketitle

\section*{Nomenclature}
\textbf{Indices}
\begin{description}
\item[$t$]~Time slot index.
\end{description}

\textbf{Constants}
\begin{description}
\item[$N$]~Total number of time slots.
\item[$\tau$]~Duration of a time slot (hour).
\item[$\theta_{\text{pv}}$]~The PV generation efficiency.
\item[$C_{\text{pv}}$]~Total radiation area of solar panels ($m^2$).
\item[$\varepsilon$]~Factor of inertia.
\item[$\eta$]~Thermal conversion efficiency (heating).
\item[$A$]~Overall thermal conductivity (kW$/^\circ F$).
\item[$\omega$]~Time constant of system (hour).
\item[$e^{\max}$]~Power rating of an HVAC system (kW).
\item[$T^{\min}$]~Lower bound of comfort range ($^\circ C$).
\item[$T^{\max}$]~Upper bound of comfort range ($^\circ C$).
\item[$T^{\text{outmin}}$]~Minimum outdoor temperature ($^\circ C$).
\item[$T^{\text{outmax}}$]~Maximum outdoor temperature ($^\circ C$).
\item[$v^{\max}$]~Maximum charging power of the EV (kW).
\item[$D^{\max}$]~Maximum queueing delay of the queue $Q_t$ (hour).
\item[$R$]~Tolerant EV charging delay (hour)
\item[$u^{\text{cmax}}$]~Maximum ESS charging power (kW).
\item[$u^{\text{dmax}}$]~Maximum ESS discharging power (kW).
\item[$\gamma$]~Thermal cost coefficient (RMB$/(^oF)^2$).
\end{description}

\textbf{Variables}
\begin{description}
\item[$r_t$]~Generation output of PV panels (kW).
\item[$\rho_t$]~Solar radiation intensity (W$/m^2$).
\item[$T_t$]~Indoor temperature ($^\circ C$).
\item[$T_t^{\text{out}}$]~Outdoor temperature ($^\circ C$).
\item[$e_t$]~HVAC input power at slot $t$ (kW).
\item[$Q_t$]~EV energy queue (kW).
\item[$x_{t}$]~Service rate of the queue $Q_t$ (kW).
\item[$a_{t}$]~Arrival rate of the queue $Q_t$ (kW).
\item[$G_t$]~Stored energy level of the ESS (kWh).
\item[$g_t$]~Purchasing/selling power of a smart home (kW).
\item[$B_t$]~Buying electricity price (RMB/kWh).
\item[$S_t$]~Selling electricity price (RMB/kWh).
\item[$y_t$]~Charging or discharging power of the ESS (kW).
\item[$T^{\text{ref}}_{t+1}$]~The most comfortable temperature ($^\circ C$).
\item[$\pi_{t+1}$]~Home occupancy state at slot $t+1$.
\item[$\Phi_{1,t}$]~Energy cost (RMB).
\item[$\Phi_{2,t}$]~Thermal discomfort cost at slot $t+1$ (RMB).
\item[$H_{t}$]~Virtual queue related to indoor temperature ($^\circ F$).
\item[$Z_{t}$]~Virtual queue related to EV charging delay (slots).
\item[$K_{t}$]~Virtual queue related to ESS control (kWh).
\item[$L_{t}$]~Lyapunov function.
\end{description}

\section{Introduction}\label{s1}

As a next-generation power system, smart grid is typified by an increased use of information and communications technology (ICT) in the generation, transmission, distribution, and consumption of electrical energy. In smart grid, there are two-way flows of electricity and information. As far as two-way information flow is concerned, consumers and utilities could exchange real-time information (e.g., electricity prices, power usages) through smart meters. Consequently, some energy management schemes could be developed to save energy cost for consumers by exploiting the temporal diversity of electricity prices\cite{Lei2012,LiangYuPowerOutage2015,LiangArxiv2016}. As large electricity consumers in power grids, residential homes account for 30\%-40\% of the total electricity consumption in a country (e.g., about 39\% in U.S.\cite{Book2011}). Therefore, it is of particular importance to carry out the design of efficient energy management for residential homes. In this paper, we mainly focus on the energy management of smart homes, which are evolved from traditional homes by adopting three components, namely the internal networks, intelligent controls, and home automations\cite{Wu2017}.

In a smart home, there are lots of appliances. In general, such appliances could be divided into two types, i.e., inflexible loads (e.g., lights, computers, and televisions) and flexible loads (e.g., heating, ventilation, and air conditioning (HVAC) systems, electric water heaters, electric vehicles (EVs), and washing machines). In this paper, we mainly focus on the scheduling of an HVAC system and an EV in a smart home, since HVAC systems account for about 50\% electricity consumption of a smart home\cite{Book2011} and EV charging task represents one of the most flexible loads (i.e., deferrable and interruptible\cite{Liang2016TSG}). As a result, the temporal price diversity could be utilized to save energy cost. In addition, distributed generation and energy storage system are also considered in the smart home. The purpose of this paper is to minimize the sum of energy cost associated with appliances and thermal discomfort cost related to occupants in a long-term time horizon with the consideration of uncertainties in electricity price, outdoor temperature, renewable generation output, electrical demand, the most comfortable temperature level, and home occupancy state. To achieve the above aim, we first formulate a problem of minimizing the time average expected total cost for the sustainable smart home with an HVAC load. Since there are time-coupling constraints and the future system parameters are unknown, it is challenging to solve the formulated problem.

Typically, the framework of Lyapunov optimization techniques (LOT)\cite{Neely2010} could be adopted to solve a time average optimization problem and an online energy management algorithm can be designed\cite{Guo2013,Fan2016}. Existing Lyapunov-based energy management algorithms intend to buffer the power demand requests of flexible loads in queues when electricity prices are high and to serve the stored requests when electricity prices are low. Different from flexible loads with specified energy/power demands (e.g., EV), an HVAC system has unknown power demand that is related to many factors, such as the most comfortable temperature level decided by occupants, the lower and upper bounds of indoor temperature, home occupancy state, and outdoor temperature. Therefore, existing Lyapunov-based energy management algorithms could not be applied to our problem directly.

To avoid knowing about an HVAC power demand when using the LOT framework, we construct a virtual queue associated with indoor temperature. By stabilizing all queues associated with indoor temperature, EV charging, and energy storage device, we design a Lyapunov-based energy management algorithm without predicting any system parameters and knowing HVAC power demand.

The main contributions of this paper are summarized as follows,
\begin{itemize}
  \item We formulate a time average expected total cost minimization for a smart home with an HVAC load considering uncertainties in electricity price, outdoor temperature, renewable generation output, electrical demand, the most comfortable temperature level, and home occupancy state.
  \item We propose an online energy management algorithm for the formulated problem based on the LOT framework without predicting any system parameters and knowing the HVAC power demand. Moreover, we theoretically analyze the feasibility and performance guarantee of the proposed algorithm.
  \item Extensive simulation results based on real-world traces show that the proposed algorithm can reduce energy cost effectively with small sacrifice in thermal comfort.
\end{itemize}

The remaining of this paper is organized as follows. Section \ref{s2} gives the literature review. In Section \ref{s3}, we describe the system model and problem formulation. Then, we propose an online algorithm to solve the formulated problem in Section \ref{s4}. In Section \ref{s5}, we conduct extensive simulations. Finally, conclusions and future work are provided in Section \ref{s6}.

\section{Literature Review}\label{s2}
There have been many studies on investigating energy management for smart homes. In \cite{Tsui2012}, Tsui \emph{et al.} proposed a convex optimization framework with $L_1$ regularization for the home energy management, which can handle appliances with ON and OFF operational statuses. In \cite{Liu2016}, Liu \emph{et al.} investigated the transformation from single-time scale into multi-time scale to reduce the computational complexity for the optimization of home energy management. In \cite{Huang2016}, Huang \emph{et al.} formulated a chance-constrained programming optimization problem to minimize energy cost of appliances considering uncertainties in a smart home. In \cite{Keerthisinghe2016}, Keerthisinghe \emph{et al.} proposed a scheme to schedule distributed energy resources in a smart home using an approximate dynamic programming with temporal difference learning. In \cite{Zhang2016}, Zhang \emph{et al.} developed a learning-based demand response strategy for an HVAC system in a smart home to minimize energy cost without affecting customer's comfort of living. In \cite{Hansen2016}, Hansen \emph{et al.} proposed a partially observable Markov decision process approach to minimize the energy cost in a smart home. In \cite{Basit2017}, Basit \emph{et al.} designed a scheme based on Dijkstra algorithm to minimize the energy cost of all devices in a home. In \cite{Deng2016}, Deng \emph{et al.} proposed a temporally-decoupled algorithm to control the indoor temperature of smart buildings with next-hour electricity price. Though some positive results have been obtained in aforementioned research efforts, they either implicitly/explicitly assume that future system parameters could be forecasted perfectly or known exactly, or require parameter forecasting.

To avoid the forecasting of system parameters, some online energy management methods have been developed based on the LOT framework\cite{Guo2013,Yang2015,Huang2014,Li2015,Shi2017}. In \cite{Guo2013}, Guo \emph{et al.} proposed a Lyapunov-based cost minimization algorithm for multiple residential households in a smart neighborhood. In \cite{Li2015}, Li \emph{et al.} investigated the joint energy storage management and load scheduling at a residential site with renewable integration and designed a real-time solution. In \cite{Yang2015}, Yang \emph{et al.} designed a cost-effective and privacy-preserving energy management strategy for smart meters by using a battery. In \cite{Huang2014}, Huang \emph{et al.} presented an adaptive electricity scheduling algorithm to minimize the microgrid operation cost with the consideration of quality-of-service in electricity. In \cite{Shi2017}, Shi \emph{et al.} proposed a real-time energy management strategy in microgrids considering physical constraints associated with the power distribution network. However, the above-mentioned works do not consider the online energy management for a smart home considering an HVAC system.

In \cite{Fan2016}, Fan \emph{et al.} investigated the online energy management problem for a smart home with an HVAC load based on the LOT framework. Specifically, this paper intends to minimize energy cost by buffering the power demand requests of appliances in queues when electricity prices are high and serving requests when electricity prices are low. However, different from loads with specific energy/power demands (e.g., EV), an HVAC load has unknown power demand in each time slot that is related to many factors, such as the most comfortable temperature level decided by occupants, the lower and upper bounds of indoor temperature, room occupancy state, and outdoor temperature. Thus, the HVAC power demand is randomly generated in \cite{Fan2016} and can not reflect the true demand of the HVAC system. Though the Lyapunov optimization technique is also adopted to design online energy management strategy for a smart home with an HVAC system, this paper has several aspects different from \cite{Fan2016}: (1) by introducing a virtual queue associated with indoor temperature, our proposed algorithm operates without knowing the HVAC power demand in each time slot; (2) we jointly consider the minimization of energy cost and thermal discomfort cost; (3) random home occupancy, energy storage system, and selling electricity are jointly considered; (4) all system control parameters which affect algorithmic feasibility are explicitly derived.

\section{System Model and Problem Formulation}\label{s3}

\begin{figure}[!htb]
\centering
\includegraphics[scale=0.58]{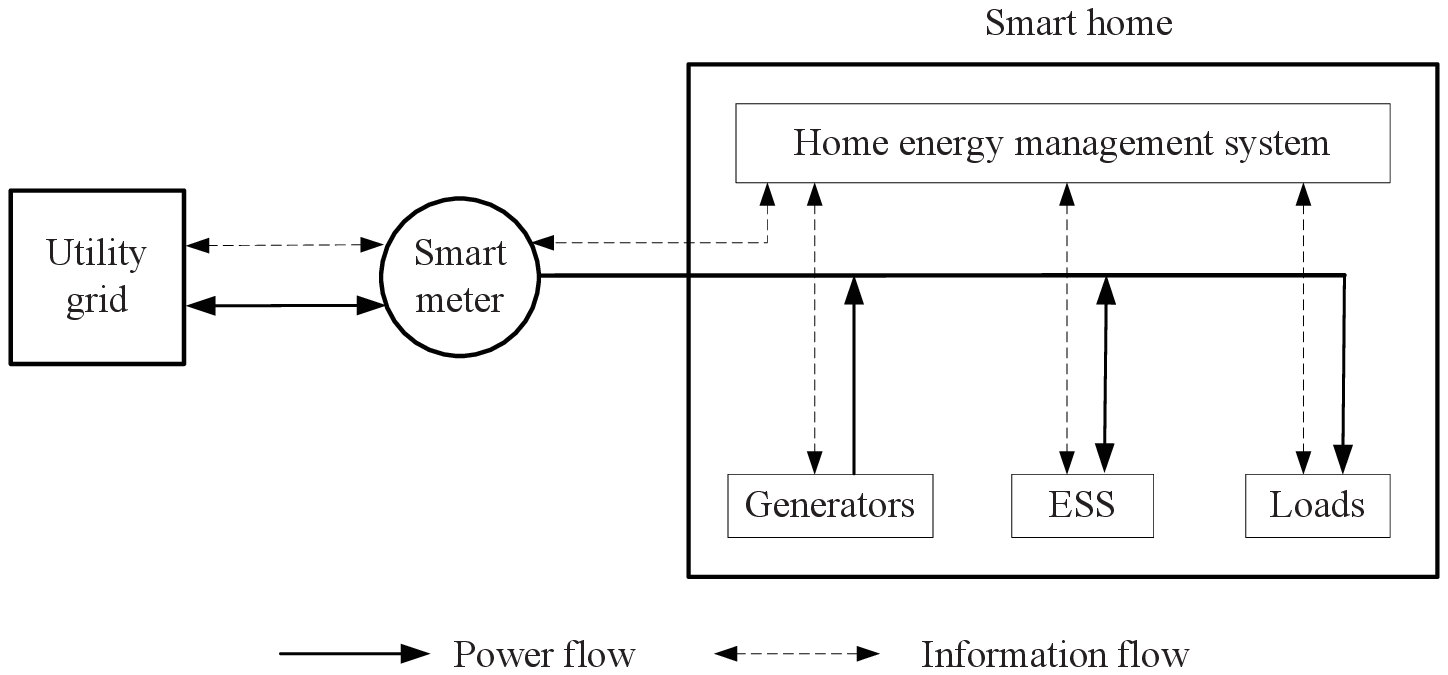}
\caption{Illustration of a smart home.}\label{fig_1}
\end{figure}

The smart home considered in this paper is shown in Fig.~\ref{fig_1}, where there are two-way communications between the smart home and the utility grid. Thus, real-time electricity price and power usage could be known by the smart home and the utility grid, respectively. In addition, the smart home could buy electricity from the utility grid and sell electricity back to the utility grid. In the smart home, some devices are connected to a low-voltage DC/AC bus for energy exchange, such as distributed generators (e.g., wind turbines or solar panels), energy storage system (ESS), home energy management system (HEMS), and loads. As the central controller of the smart home, HEMS manages the way of energy generation, storage, and consumption. Specifically, through two-way information flow, HEMS could collect system states (e.g., renewable generation output, electricity price, outdoor temperature, energy demand, and home occupancy state) and send control signals to controllable devices. As far as loads are concerned, we mainly consider flexible loads such as HVAC and EV, while the demands of other loads are satisfied by HEMS instantly and not considered in the optimization problem. In addition, HEMS operates in slotted time, i.e., $t\in [0,T]$, where $T$ is the total number of time slots. Moreover, the duration of a time slot $\tau$ is normalized to a unit time so that power and energy could be used equivalently.

\subsection{Renewable energy model}
We consider the solar energy supply in the smart home. Let $r_t$ be the maximum generation output of photovoltaic (PV) solar panels at slot $t$. Then, $r_t$ could be estimated by the following model\cite{Li2016}, i.e.,
\begin{align} \label{f_0}
r_t=\theta_{\text{pv}} C_{\text{pv}} \rho_t,~\forall~t
\end{align}
where $\theta_{\text{pv}}$ denotes the PV generation efficiency, $C_{\text{pv}}$ is the total radiation area of solar panels (in $m^2$), and $\rho_t$ represents the solar radiation (in $W/m^2$).

\subsection{Load model}

\subsubsection{HVAC model}
Generally, an HVAC system has two kinds of operational modes, i.e., heating and cooling. In this paper, we mainly focus on the heating mode in the winter and our designed algorithm could be extended to accommodate the cooling mode in the summer by taking the changes of equations \eqref{f_1} and \eqref{f_17}-(19) into consideration, e.g., the second ``+" in \eqref{f_1} should be ``-" when the cooling mode is considered.

According to \cite{Thatte2012}, the indoor temperature dynamics caused by an HVAC system could be obtained as follows
\begin{align} \label{f_1}
T_{t+1}=\varepsilon T_{t}+(1-\varepsilon)(T_t^{\text{out}}+\frac{\eta}{A}e_t),~\forall~t
\end{align}
where $T_t$ and $T_t^{\text{out}}$ denote the indoor temperature and outdoor temperature, respectively; $\eta$ is thermal conversion efficiency, and $A$ is the overall thermal conductivity in $kW/^\circ F$; $\varepsilon=e^{-\tau /\omega}$; $\omega$ is the system time constant.

In this paper, we consider an HVAC system with inverter, i.e., the HVAC system can adjust its input power $e_t$ continuously\cite{Song2017}. Let $e^{\max}$ be the rating power of the HVAC system, we have
\begin{align} \label{f_2}
0\leq e_t\leq e^{\max},~\forall~t.
\end{align}

For one person in the smart home, he/she would feel comfortable when the indoor temperature varies within a range, e.g., $20^\circ C\sim 25^\circ C$. Thus, we have the following constraints,
\begin{align} \label{f_3}
T^{\min}\leq T_t\leq T^{\max},~\forall~t,
\end{align}
where $T^{\min}$ and $T^{\max}$ are lower and upper bounds of the comfort range, respectively.

\subsubsection{EV charging model}
When an EV is connected to the DC/AC bus in the smart home, EV should send a charging request to the HEMS. To be specific, the charging request is represented by a 3-tuple ($s$,~$c$,~$E$), where $s$, $c$ and $E$ denote the starting time, completion time and energy demand, respectively. To fully utilize the temporal diversity of dynamic prices, the energy demand of the EV should be satisfied intelligently, i.e., executing charging when electricity prices are low and deferring charging process when electricity prices are high. To intelligently satisfy the energy demand of the EV without violating the completion time, we adopt an energy queue $Q_t$ as follows,
\begin{align}\label{f_4}
Q_{t+1}=\max[Q_{t}-x_{t},0]+a_{t},~\forall t,
 \end{align}
where $x_{t}$ and $a_{t}$ are service and arrival processes of the energy queue, respectively. Since the LOT framework could transform a long-term optimization problem into many online subproblems through the queue stability control, introducing the energy queue $Q_t$ contributes to online scheduling of EV charging as in \cite{LiangIoT2016}.

Denote the maximum value of $x_{t}$ by $x^{\max}$, where $x^{\max}\geq a^{\max}$ ($a^{\max}=\max_t a_{t}$) so that it is always possible to make the queue $Q_{t}$ stable. Note that there is no need to serve the energy demand that is greater than $Q_{t}$, we have
\begin{align}\label{f_5}
0\leq x_{t}\leq \min\{x^{\max},Q_{t}\},~\forall t.
\end{align}

Since the charging power of an EV is limited, the EV can add at most an energy demand $v^{\max}$ to $Q_{t}$ in a time slot, where $v^{\max}$ denotes the maximum charging power of the EV. When $E>v^{\max}$, multiple time slots are needed to finish the submission of total energy demand $E$. Similar to \cite{LiangIoT2016}, the EV submits the energy demand at slot $t$ according to the following equation,
\begin{align}\label{f_6}
a_{t}  = \left\{ \begin{array}{l}
 v^{\max } ,
\begin{array}{*{20}c}
   {}  \\
\end{array}\begin{array}{*{20}c}
   {}  \\
\end{array}\begin{array}{*{20}c}
   {}  \\
\end{array}s  \le t < s  + \kappa;  \\
 E  - \kappa v^{\max } ,\begin{array}{*{20}c}
   {}  \\
\end{array}t = s  + \kappa ;  \\
 0,\begin{array}{*{20}c}
   {}  \\
\end{array}\begin{array}{*{20}c}
   {\begin{array}{*{20}c}
   {}  \\
\end{array}}  \\
\end{array}\begin{array}{*{20}c}
   {\begin{array}{*{20}c}
   {}  \\
\end{array}}  \\
\end{array}\text{otherwise}, \\
 \end{array} \right.
\end{align}
where $\kappa=\left\lfloor {\frac{{e }}{{v^{\max } }}} \right\rfloor$.

To ensure that the average length of $Q_t$ is finite, we have the following constraint, i.e.,
\begin{align}\label{f_7}
\mathop {\lim\sup}\limits_{T \to \infty}\frac{1}{T}\sum\nolimits_{t=0}^{T-1} \mathbb{E}\{Q_{t}\}<\infty.
\end{align}

Since \eqref{f_7} is not enough to ensure that the charging completion time is not violated, we adopt the following constraint,
\begin{align} \label{f_8}
D^{\max}\leq R,
\end{align}
where $D^{\max}$ is the maximum queueing delay of the queue $Q_{t}$; $R$ denotes the tolerant EV charging service delay, which is equal to $c-s-\kappa $.

\subsection{ESS model}
Let $y_t$ be the charging or discharging power of the ESS at slot $t$. Then, we have
\begin{align}\label{f_9}
-u^{\text{dmax}}\leq y_t\leq u^{\text{cmax}},~\forall t,
\end{align}
where $u^{\text{cmax}}>0$ and $u^{\text{dmax}}>0$ are maximum charging power and discharging power, respectively. Following the ESS models in \cite{Guo2013},\cite{Huang2014},\cite{Guan2010}, the dynamics of energy levels in the ESS could be expressed by
\begin{align}\label{f_10}
G_{t+1}=G_{t}+ y_t,~\forall t,
\end{align}
where $G_{t}$ represents the stored energy of the ESS at time slot $t$. Though the ESS model with perfect charging and discharging efficiency parameters is considered in this paper, the proposed Lyapunov-based algorithm can be extended to incorporate more complex ESS models, please see \cite{LiangAccess2016} for details.

Since the energy level should fluctuate within a certain range, we have
\begin{align}\label{f_11}
G^{\min}\leq G_{t}\leq G^{\max},~\forall t,
\end{align}
where $G^{\max}$ and $G^{\min}$ denote the maximum and the minimum capacity of the ESS, respectively.

\subsection{Power balancing}
Let $g_t$ be the energy transaction between the smart home and the main grid at slot $t$. Specifically, $g_t>0$ means electricity purchasing while $g_t< 0$ means electricity selling. Then, according to the real-time power balancing, we have
\begin{align} \label{f_12}
g_t+r_t=e_t+x_t+y_t,~\forall~t.
\end{align}

\subsection{Problem formulation}

With the above models, the energy cost due to the electricity selling or buying is given by
\begin{align} \label{f_13}
\Phi_{1,t}=\Bigg(\frac{B_t-S_t}{2}|g_{t}|+\frac{B_t+S_t}{2}g_{t}\Bigg),
\end{align}
where $B_t\in [B^{\min},~B^{\max}]$ and $S_t \in [S^{\min},~S^{\max}]$ denote the price of buying and selling electricity at slot $t$, respectively. We assume that selling prices are not higher than purchasing prices, i.e., $B_t\geq S_t$ for all $t$. In other words, the smart home cannot make profit by greedily purchasing energy from the utility grid and then selling it back to the utility grid at a higher price simultaneously. Such assumption is commonly made in existing works\cite{Huang2014}\cite{Zhangyu2013}. In addition, the intuition behind \eqref{f_13} is that just a variable $g_t$ is needed to reflect the electricity purchasing or selling. For example, when $g_t\leq 0$, $\Phi_{1,t}=S_tg_{t}$. For the case $g_t>0$, $\Phi_{1,t}=B_tg_{t}$.

Similar to \cite{Constantopoulos1991}, we model the thermal discomfort cost of occupants at slot $t$ by
\begin{align} \label{f_15}
\Phi_{2,t}=\gamma \pi_{t+1}(T_{t+1}-T^{\text{ref}}_{t+1})^2,
\end{align}
where $\gamma$ is the cost coefficient with unit $\$/(^oF)^2$, which reflects the relative importance of discomfort cost with respect to energy cost; $T^{\text{ref}}_{t+1}$ denotes the most comfortable level for occupants in slot $t+1$ (e.g., 22.5$^oC$), and its value could be decided by occupants at slot $t$. Binary variable $\pi_{t+1}$ represents the home occupancy state at time slot $t+1$ (``1" denotes occupancy and ``0" denotes vacancy). When $\pi_{t+1}=0$, $\Phi_{2,t}$ would be zero since there is no occupant at home. The value of $\pi_{t+1}$ could be decided by the last occupant, who is going to leave the home at slot $t$. If no human participation is expected, smart devices with sensors could be deployed to implement behavior awareness (e.g., leave home) and execute the corresponding operations\cite{Chen2017}.

With above-mentioned models, we formulate a problem to minimize the sum of energy cost and thermal discomfort cost as follows,
\begin{subequations}\label{f_16}
\begin{align}
(\textbf{P1})~&\min~\mathop {\lim\sup}\limits_{N \to \infty}\frac{1}{N-1}\sum\limits_{t=0}^{N-2} \mathbb{E}\{\Phi_{1,t}+\Phi_{2,t}\}  \\
s.t.&~(2)-(6),(8)-(13),
\end{align}
\end{subequations}
where $\mathbb{E}$ denotes the expectation operator, which acts on random purchasing/selling electricity prices $B_t$/$S_t$, outdoor temperatures $T_t^{\text{out}}$, renewable generation outputs $r_t$, EV electrical demand $a_t$, the most comfortable temperature level $T_{t+1}^{\text{ref}}$, and home occupancy state $T_{t+1}^{\text{ref}}$; the decision variables of \textbf{P1} are $e_t$, $x_t$, $y_t$ and $g_{t}$.

\section{Algorithm Design}\label{s4}
\subsection{The proposed online algorithm}
There are two challenges involved in solving \textbf{P1}. Firstly, the constraints \eqref{f_1},~\eqref{f_4},~\eqref{f_10} introduce time couplings, which means that the current decision would affect future decisions. Secondly, future parameters are unknown, e.g., electricity prices and outdoor temperatures. To handle the ``time-coupling" property, typical methods are based on dynamic programming\cite{Bertsekas2000}, which suffers from ``the curse of dimensionality" problem. Recently, the LOT framework was often adopted to deal with the above challenges\cite{Guo2013}\cite{Fan2016}. Existing Lyapunov-based energy management algorithms intend to buffer the power demand requests of appliances (e.g., EV) in queues and to serve such requests when electricity prices are low. Different from EVs with specified energy/power demands, an HVAC load has unknown power demand in each time slot that is related to many factors, such as the most comfortable temperature level decided by occupants, the lower and upper bounds of indoor temperature, room occupancy state, and outdoor temperature. Thus, we need to redesign an algorithm to deal with the HVAC load. The key idea of the proposed algorithm is summarized as follows:
\begin{itemize}
  \item Constructing virtual queues associated with indoor temperature, EV charging delay and ESS.
  \item Obtaining the \emph{drift-plus-penalty} term according to the LOT framework.
  \item Minimizing the upper bound given in the right-hand-side of the \emph{drift-plus-penalty} term.
\end{itemize}

Based on the above idea, we can propose an online energy management algorithm without predicting any system parameters and knowing HVAC power demand in each time slot. Note that the purpose of constructing virtual queues is to guarantee the feasibility of constraints \eqref{f_3},~\eqref{f_8}, and \eqref{f_11}. By stabilizing such queues, the proposed algorithm could operate without violating the constraints \eqref{f_3},~\eqref{f_8}, and \eqref{f_11}. Specific proof can be found in Theorems 1-3.

To begin with, three mild assumptions are made about system parameters so that the system is controllable, i.e.,
\begin{align}\label{f_17}
&~~~~~~~~~~~~~~~T^{\text{outmax}}\leq T^{\max}, \\
&~~~~~~~~~~\frac{\eta}{A}e^{\max}+T^{\text{outmin}}\geq T^{\min},\\
&~~~~~~~~~~~~~~T^{\max}-T^{\min}>\psi,
\end{align}
where $T^{\text{out}}\in [T^{\text{outmin}},~T^{\text{outmax}}]$, $\psi=(1-\varepsilon)(T^{\text{outmax}}-T^{\text{outmin}}+\frac{\eta}{A}e^{\max})$. Note that the first assumption is very common for the heating mode in the winter since the highest temperature in the winter is always less than the comfortable temperature levels (e.g., Fig.~\ref{fig_2}(a) shows that $T^{\text{outmax}}$ is about $10^oC$, while $T^{\max}$ is about 25$^oC$.). In addition, the second assumption simply implies that the temperature decay can be compensated by injecting the maximum power of the HVAC system (this is required by any an HVAC system). The last assumption could also be satisfied easily in practice, e.g., when we set the parameters as follows\cite{Constantopoulos1991}\cite{Deng2016}: $T^{\max}=23.5^\circ C$, $T^{\min}=20^\circ C$, $\varepsilon=0.96$, $T^{\text{outmax}}-T^{\text{outmin}}=10^\circ C$, $\eta=1$, $A=0.14kW/^\circ F$, $e^{\max}=10kW$, we have $\psi=4.8571^oF<T^{\max}-T^{\min}=6.3^oF$. The intuition behind (19) is that the control parameter $V_1^{\max}$ in (33) should be greater than zero, i.e., $d>0$.

\subsubsection{Constructing virtual queues}
To guarantee the feasibility of \eqref{f_3}, we define a virtual queue as a shifted version of indoor temperature $T_t$ as follows,
\begin{align} \label{f_18}
H_{t}=T_{t}+\Gamma,
\end{align}
where $\Gamma$ is a constant, which is specified in the Theorem 1 of Section \ref{s4}-B. Continually, the dynamics of $H_t$ could be obtained below,
\begin{align} \label{f_19}
H_{t+1}=\varepsilon H_t+(1-\varepsilon)(\Gamma+T_t^{\text{out}}+\frac{\eta}{A}e_t),
\end{align}
where the above equation could be obtained by integrating \eqref{f_1} with \eqref{f_18}.

In addition, to satisfy the requirement of \eqref{f_8}, we define a delay-aware virtual queue $Z_t$ as follows,
\begin{align} \label{f_20}
Z_{t + 1}  = \left\{ \begin{array}{l}
 [Z_{t}  - x_{t}  + \xi]^ +, Q_{t}> x_{t}, \\
 0,~~\begin{array}{*{20}c}
   {}  \\
\end{array}\begin{array}{*{20}c}
   {}  \\
\end{array}\begin{array}{*{20}c}
   {}  \\
\end{array}\begin{array}{*{20}c}
   {}  \\
\end{array}\begin{array}{*{20}c}
   {}  \\
\end{array}Q_{t}  \le x_{t}, \\
 \end{array} \right.
\end{align}
where $[\diamond]^+\triangleq \max\{\diamond,0\}$; $\xi$ is a fixed parameter, which represents the arrival rate of the virtual queue $Z_t$ when the queue $Q_t>x_t$, while $x_t$ represents the service rate of $Z_t$. According to our previous work\cite{LiangArxiv2016}, it can be known that \eqref{f_8} could be guaranteed when the queues $Q_{t}$ and $Z_{t}$ have finite upper bounds. Moreover, the maximum queueing delay $D^{\max}=\lceil(Q^{\max}+Z^{\max})/\xi\rceil$. In next section, we will show that such upper bounds indeed exist.

To guarantee the feasibility of \eqref{f_11}, we define a virtual queue as a shifted version of ESS energy level $G_t$ as follows,
\begin{align} \label{f_21}
K_t=G_{t}+\alpha,
\end{align}
where $\alpha$ is a constant, which is specified in the Theorem 3 of Section \ref{s4}-B. Continually, the dynamics of $K_t$ is given by
\begin{align} \label{f_22}
K_{t+1}=K_t+\alpha,
\end{align}

\subsubsection{Obtaining \emph{drift-plus-penalty} term}

In addition to keeping three virtual queues stable, the actual energy queue $Q_t$ should also be stabilized so that \eqref{f_7} could be satisfied. Thus, we define a Lyapunov function below,
\begin{align} \label{f_23}
L_t=\frac{1}{2}(H_t^2+Q_t^2+Z_t^2+K_t^2).
\end{align}

Define $\boldsymbol{\Psi_t}\triangleq (H_t,Q_t,Z_t,K_t)$, the one-slot conditional Lyapunov drift could be calculated as follows,
\begin{align} \label{f_24}
    \Delta_t = \mathbb{E}\{ L_{t+1} - L_t|\boldsymbol{\Psi_t}\},
\end{align}
where the expectation is taken with respect to the randomness of electricity prices, outdoor temperatures, renewable generation output, EV charging demand, the most comfortable temperature level, and home occupancy state, as well as the chosen control decisions.

Taking \eqref{f_23} into consideration, we have
 \begin{align}\label{f_25}
 L_{t+1} - L_t= \big(\varphi_H+\varphi_Q+\varphi_Z+\varphi_K\big),
 \end{align}
where $\varphi_H=\frac{1}{2}(H_{t+1}^2-H_t^2)$,~$\varphi_Q=\frac{1}{2}(Q_{t+1}^2-Q_t^2)$,~$\varphi_Z=\frac{1}{2}(Z_{t+1}^2-Z_t^2)$,~$\varphi_K=\frac{1}{2}(K_{t+1}^2-K_t^2)$. Specifically, $\varphi_H$, $\varphi_Q$, $\varphi_Z$, and $\varphi_K$ have the following upper bounds, i.e.,
 \begin{align}\label{f_26}
\varphi_H = &\frac{1}{2}(H_{t + 1}^2 - H_t^2) \nonumber \\
&< {\Omega_0} + \varepsilon (1 - \varepsilon ){H_t}(\Gamma  + T_t^{\text{out}} + \frac{\eta }{A}{e_t}), \\
\varphi_Q= &\frac{1}{2}(Q_{t + 1}^2 - Q_t^2) < {\Omega _1} + {Q_t}({a_t} - {x_t}),\\
\varphi_Z= & \frac{1}{2}(Z_{t + 1}^2 - Z_t^2) < {\Omega _2} + Z_t(\xi-{x_t}), \\
\varphi_K= & \frac{1}{2}(K_{t+1}^2 - K_t^2) <\Omega_3 + K_ty_t,
 \end{align}
 where $\Omega_0=\frac{(1-\varepsilon)^2}{2}\max\Big((\Gamma+T^{\text{outmin}})^2,(\Gamma+T^{\text{outmax}}+\frac{\eta}{A}e^{\max})^2\Big)$, $\Omega_1=\frac{(x^{\max})^2+(a^{\max})^2}{2}$, $\Omega_2=\frac{1}{2}\max(\xi^2,(x^{\max})^2)$, $\Omega_3=\frac{(\max(u^{\text{cmax}},u^{\text{dmax}}))^2}{2}$.

By adding a function of the expected total cost over one slot to \eqref{f_24}, we can obtain the \emph{drift-plus-penalty} term as follows,
\begin{align}\label{f_27}
&\Delta Y_t=\Delta_t + V\mathbb{E}\{ {\Phi_{1,t}+\Phi_{2,t}|\boldsymbol{\Psi_t}} \}\nonumber\\
\leq& \sum\nolimits_{l=1}^4\Omega_l+\mathbb{E}\{K_ty_t-(Q_t+Z_t)x_t|\boldsymbol{\Psi_t}\}\nonumber \\
&+\mathbb{E}\{\varepsilon (1 - \varepsilon ){H_t}(\Gamma  + T_t^{\text{out}} + \frac{\eta }{A}{e_t})|\boldsymbol{\Psi_t}\}\nonumber \\
&+V\mathbb{E}\{ {\Phi_{1,t}+\Phi_{2,t}|\boldsymbol{\Psi_t}} \},
\end{align}
where $V$ is a weight parameter that implements a tradeoff between queue stability and total cost reduction.

\subsubsection{Minimizing the upper bound}
Since the main principle of the Lyapunov-based algorithm is to choose control actions that minimize the upper bound given in
the right-hand-side of the \emph{drift-plus-penalty} term. Then, the proposed algorithm could be described by the Algorithm 1, where \textbf{P2} is a convex optimization problem with four variables and its solution could be solved efficiently using available convex methods (e.g., interior point methods) or tools (e.g., CVX). In addition, the value of $T_{t+1}$ with $e_t=0$ is described by $T_{t+1}|_{e_{t}=0}$ for brevity. The lines 6-7 denote that the HVAC power input would be zero if the home is not occupied in the next time slot and $T_{t+1}|_{e_{t}=0}$ is still greater than $T^{\min}$, which contributes to saving energy cost without affecting the thermal comfort of occupants.

\begin{algorithm}[h]
\caption{: Home Energy Management Algorithm}
\begin{algorithmic}[1]
\STATE \textbf{For} each slot $t$ \textbf{do}\\
\STATE At the beginning of slot $t$, observe $\boldsymbol{\Psi_t}$,$T_t^{\text{out}}$, $B_t$, $S_t$, $r_t$, $a_t$, $T_{t+1}^{\text{ref}}$, $\pi_{t+1}$; \\
\STATE Choose $g_t$, $e_t$, $x_t$, and $y_t$ as the solution to \textbf{P2}:
\STATE (\textbf{P2})~$\min~K_ty_t-(Q_t+Z_t)x_t+\varepsilon (1 - \varepsilon ){H_t}(\Gamma  + T_t^{\text{out}} + \frac{\eta }{A}{e_t})+V(\Phi_{1,t}+\Phi_{2,t})$  \\
\STATE ~~~~~~s.t.~(3),~(6),~(10),~(13) \\
\STATE \textbf{If} $\pi_{t+1}=0$ and $T_{t+1}|_{e_{t}=0}\geq T^{\min}$ \\
\STATE ~~~~~~$e_{t}=0$,~$g_t=x_t+y_t-r_t$
\STATE \textbf{End}
\STATE Update $H_t$, $Q_t$, $Z_t$ and $K_t$ according to \eqref{f_19},\eqref{f_4},\eqref{f_20},~\eqref{f_22};
\STATE \textbf{End}
\end{algorithmic}
\end{algorithm}

Note that updating $H_t$, $Q_t$, and $K_t$ according to \eqref{f_19},~\eqref{f_4},~\eqref{f_22} means that the constraints \eqref{f_1},~\eqref{f_4},~\eqref{f_10} could be satisfied by the proposed algorithm. Moreover, (3),~(6),~(10),~(13) are explicitly incorporated in \textbf{P2}. Thus, the remaining constraints (i.e., \eqref{f_3},~\eqref{f_7},~\eqref{f_8},~\eqref{f_11}) are not considered in Algorithm 1. In the next section, we will show the feasibility of the proposed algorithm for \textbf{P1} by proving that \eqref{f_3},~\eqref{f_7},~\eqref{f_8},~\eqref{f_11} could be satisfied.

\subsection{Algorithm feasibility}
Let ($e_t^*$,$x_t^*$,$y_t^*$,$g_t^*$) be the optimal solution of \textbf{P2}, we have the following three Lemmas and three Theorems, which show that the constraints \eqref{f_3},~\eqref{f_7},~\eqref{f_8},~\eqref{f_11} could be satisfied by the proposed algorithm.

\textbf{\emph{Lemma 1.}}
The optimal HVAC operation decision of the proposed algorithm has the following properties, where $b_t=2V\gamma \pi_{t+1} (1-\varepsilon)^2\eta/A(T_t^{\text{out}}-\frac{T^{\text{ref}}_{t+1}-\varepsilon T_t}{1-\varepsilon})$, $c_t=2V\gamma \pi_{t+1} (1-\varepsilon)^2\eta/A(T_t^{\text{out}}+\frac{\eta}{A}e^{\max}-\frac{T^{\text{ref}}_{t+1}-\varepsilon T_t}{1-\varepsilon})$.
\begin{enumerate}
  \item If $VS^{\min}+b_t>-\varepsilon(1-\varepsilon)H_t\frac{\eta}{A}$, $e_t=0$.
  \item If $VB^{\max}+c_t<-\varepsilon(1-\varepsilon)H_t\frac{\eta}{A}$, $e_t=e^{\max}$.
\end{enumerate}
\begin{IEEEproof}
See Appendix A.
\end{IEEEproof}

Based on Lemma 1, Theorem 1 could be derived as follows.

\textbf{\emph{Theorem 1.}}
\emph{If the initial temperature $S_0\in [T^{\min},~T^{\max}]$, the proposed algorithm with fixed parameters $V \in (0, V_1^{\max}]$ and $\Gamma \in [\Gamma^{\min},~\Gamma^{\max}]$ would offer the following guarantee, i.e., $T_t \in [T^{\min},~T^{\max}]$ for all time slots, where}
\begin{align}
&V_1^{\max}=\frac{(1-\varepsilon)\frac{\eta}{A}d}{(B^{\max}-S^{\min})+f},\\
&\Gamma^{\min}=\frac{VS^{\min}+b^{\min}}{-\varepsilon(1-\varepsilon)\frac{\eta}{A}}+\frac{h}{\varepsilon}, \\
&\Gamma^{\max}=\frac{VB^{\max}+c^{\max}}{-\varepsilon(1-\varepsilon)\frac{\eta}{A}}+\frac{m}{\varepsilon}.
\end{align}
In above formulas, $d=T^{\max}-T^{\min}-(1-\varepsilon)(T^{\text{outmax}}+\frac{\eta}{A}e^{\max}-T^{\text{outmin}})$, $f=2\gamma(1-\varepsilon)^2\eta(T^{\text{outmax}}-T^{\text{outmin}}+e^{\max}+\frac{\varepsilon(T^{\max}-T^{\min})+(T^{\text{refmax}}-T^{\text{refmin}})}{1-\varepsilon})/A$, $h=(1-\varepsilon)(T^{\text{outmax}}+\frac{\eta}{A}e^{\max})-T^{\max}$, $m=(1-\varepsilon)T^{\text{outmin}}-T^{\min}$, $b^{\min}=\min_t b_t$, $c^{\max}=\max_t c_t$, $T^{\text{refmax}}=\max_t T_t^{\text{ref}}$, $T^{\text{refmin}}=\min_t T_t^{\text{ref}}$.

\begin{IEEEproof}
See Appendix B.
\end{IEEEproof}

\textbf{\emph{Lemma 2.}}
The optimal EV charging decision of the proposed algorithm has the following properties:
\begin{enumerate}
  \item If $Q_t+Z_t<VS^{\min}$, $x_t^*=0$.
  \item If $Q_t+Z_t>VB^{\max}$, $x_t^*=\min\{x^{\max},Q_t\}$.
\end{enumerate}
\begin{IEEEproof}
See Appendix C.
\end{IEEEproof}

Based on Lemma 2, Theorem 2 could be derived as follows.

\textbf{\emph{Theorem 2}}
Suppose $x^{\max}\geq \max[a^{\max},~\xi]$. If $Q_{0}=Z_{0}=0$, the proposed online algorithm has the following properties:
\begin{enumerate}
  \item $Q_{t}$ is bounded by $Q^{\max}=VB^{\max}+a^{\max}$, $Z_t$ is bounded by $Z^{\max}=VB^{\max}+\xi$.
  \item Maximum queueing delay $D^{\max}=\left\lceil\frac{2VB^{\max}+a^{\max}+\xi}{\xi}\right\rceil.$
\end{enumerate}

\begin{IEEEproof}
See Appendix D.
\end{IEEEproof}

\textbf{\emph{Lemma 3.}}
The optimal ESS decision of the proposed algorithm has the following properties:
\begin{enumerate}
  \item If $K_t>-VS^{\min}$, we have $y_t^*\leq 0$.
  \item If $K_t<-VB^{\max}$, we have $y_t^*\geq 0$.
\end{enumerate}
\begin{IEEEproof}
See Appendix E.
\end{IEEEproof}

Based on Lemma 3, Theorem 3 could be derived as follows.

\textbf{\emph{Theorem 3.}}
If the initial energy level $G_0\in [G^{\min},~G^{\max}]$, the proposed algorithm with fixed parameters $V \in (0, V_2^{\max}]$ and $\alpha \in [\alpha^{\min},~\alpha^{\max}]$ would offer the following guarantee, i.e., $G_t \in [G^{\min},~G^{\max}]$ for all slots, where
\begin{align}
&V_2^{\max}=\frac{G^{\max}-G^{\min}-(u^{\text{cmax}}+u^{\text{dmax}})}{(B^{\max}-S^{\min})},\\
&\alpha^{\min}=-VS^{\min}+u^{\text{cmax}}-G^{\max}, \\
&\alpha^{\max}=-VB^{\max}-u^{\text{dmax}}-G^{\min},
\end{align}

\begin{IEEEproof}
See Appendix F.
\end{IEEEproof}

\textbf{\emph{Theorems 1-3}} show that the constraints \eqref{f_3},~\eqref{f_7},~\eqref{f_8},~\eqref{f_11} could be ensured under the proposed algorithm. Since other constraints are explicitly considered in Algorithm 1, we have the conclusion that the proposed algorithm is feasible to the original problem \textbf{P1}. In next section, we will use real-world traces about outdoor temperature, electricity price and PV generation to test the effectiveness of the proposed algorithm.

\subsection{Performance guarantee}
In this subsection, we analyze the performance guarantee of the proposed algorithm in Theorem 4.

\textbf{\emph{Theorem 4.}}
If purchasing/selling electricity prices $B_t$/$S_t$, outdoor temperatures $T_t^{\text{out}}$, renewable generation outputs $r_t$, EV electrical demand $a_t$, the most comfortable temperature level $T_{t+1}^{\text{ref}}$, and home occupancy state $T_{t+1}^{\text{ref}}$ are i.i.d. over slots, the proposed algorithm offers the following performance guarantee, i.e., $\mathop {\lim\sup}\limits_{N \to \infty}\frac{1}{N-1}\sum\nolimits_{t=0}^{N-2} \mathbb{E}\{\Phi_{1,t}+\Phi_{2,t}\}\leq y_1+\frac{\Theta}{V}$, where $y_1$ is the optimal objective value of \textbf{P1}.

\begin{IEEEproof}
See Appendix G.
\end{IEEEproof}
Since $\Theta$ is a complex function of $V$, the above optimality gap would not monotonically decreases with the increase of $V$. Specially, when $\varepsilon=1$, $\Theta$ becomes a constant. At this time, the performance of the proposed algorithm would be better given a greater $V$. When uncertain parameters are non i.i.d over slots, performance analysis could be conducted using ``multi-slot drift" in \cite{Neely2010}, which will be our future work.

\section{Performance Evaluation}\label{s5}
\subsection{Simulation setup}
\begin{figure}
\centering
\subfigure[Outdoor temperature]{
\begin{minipage}[b]{0.4\textwidth}
\includegraphics[width=1\textwidth]{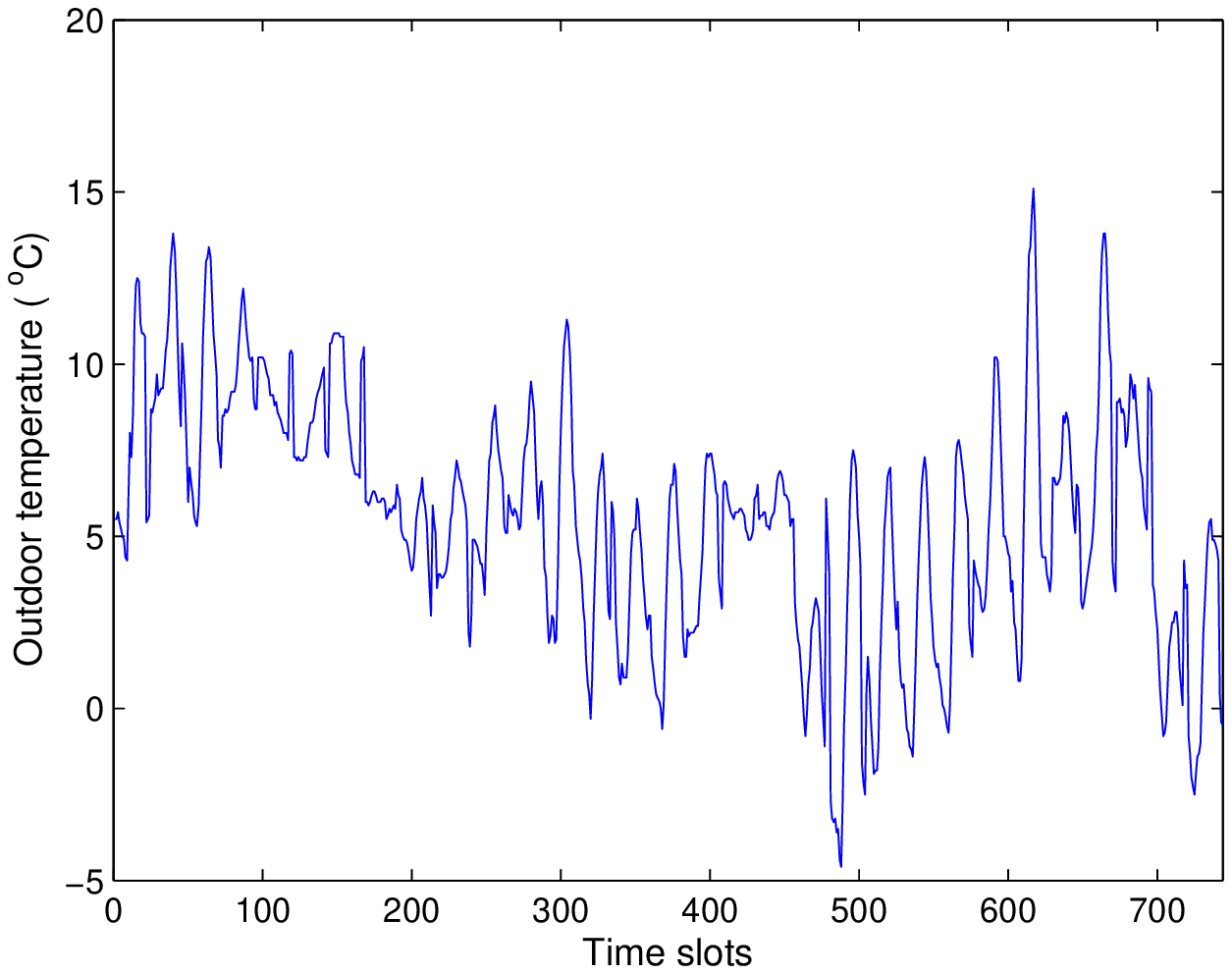}
\end{minipage}
}\\
\subfigure[Retail electricity price]{
\begin{minipage}[b]{0.4\textwidth}
\includegraphics[width=1\textwidth]{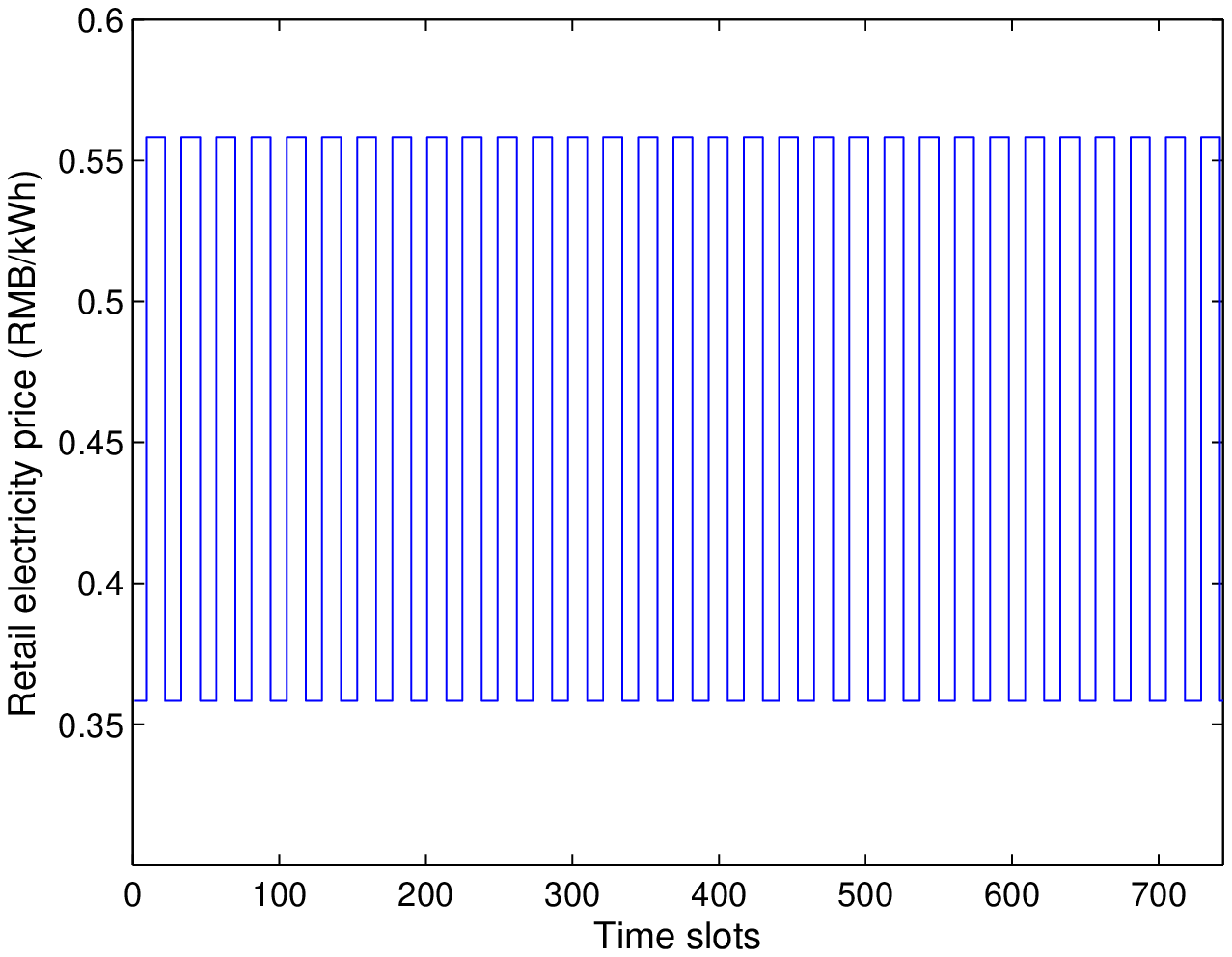}
\end{minipage}
}\\
\subfigure[Solar radiation]{
\begin{minipage}[b]{0.4\textwidth}
\includegraphics[width=1\textwidth]{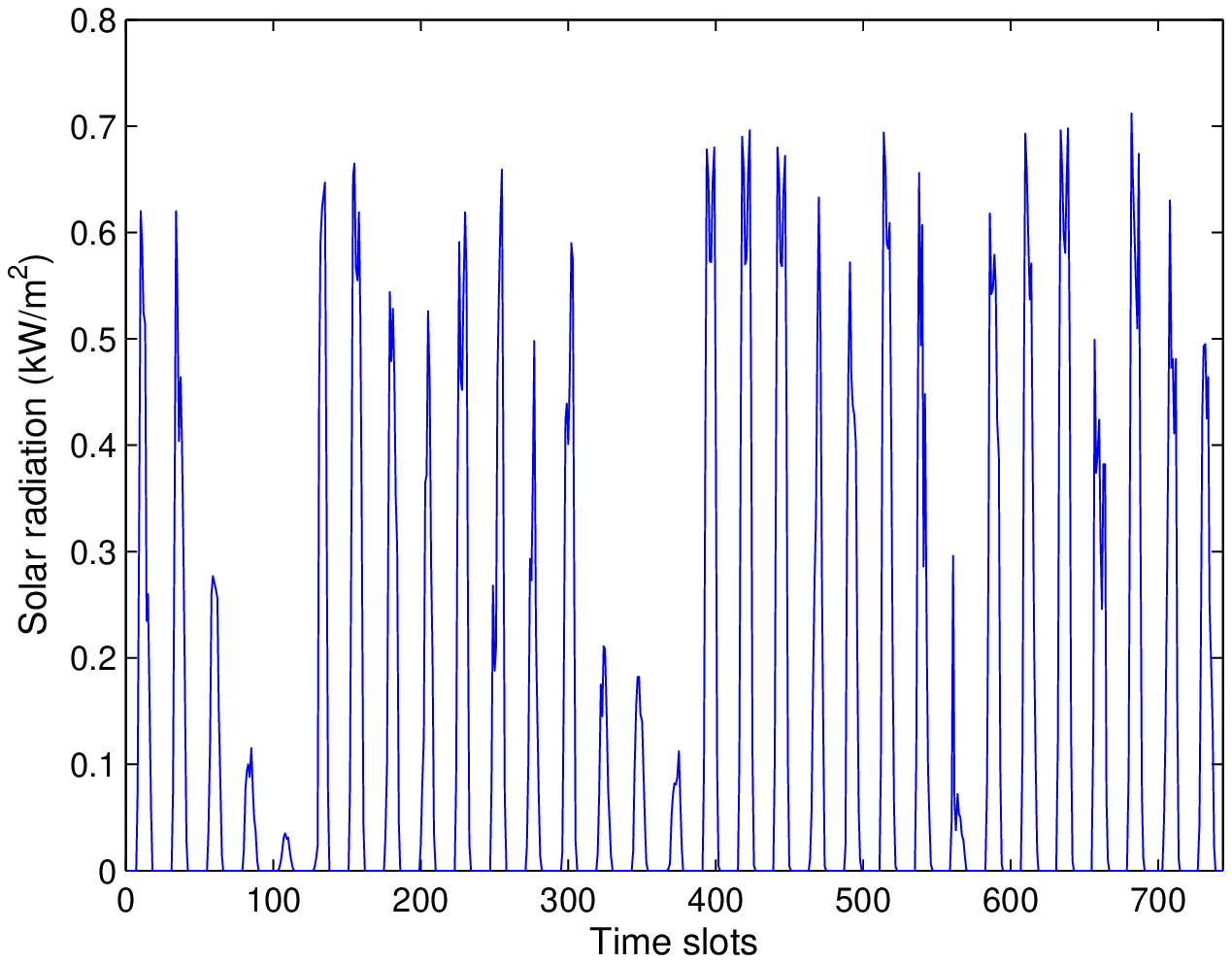}
\end{minipage}
}\\
\subfigure[Occupancy state]{
\begin{minipage}[b]{0.4\textwidth}
\includegraphics[width=1\textwidth]{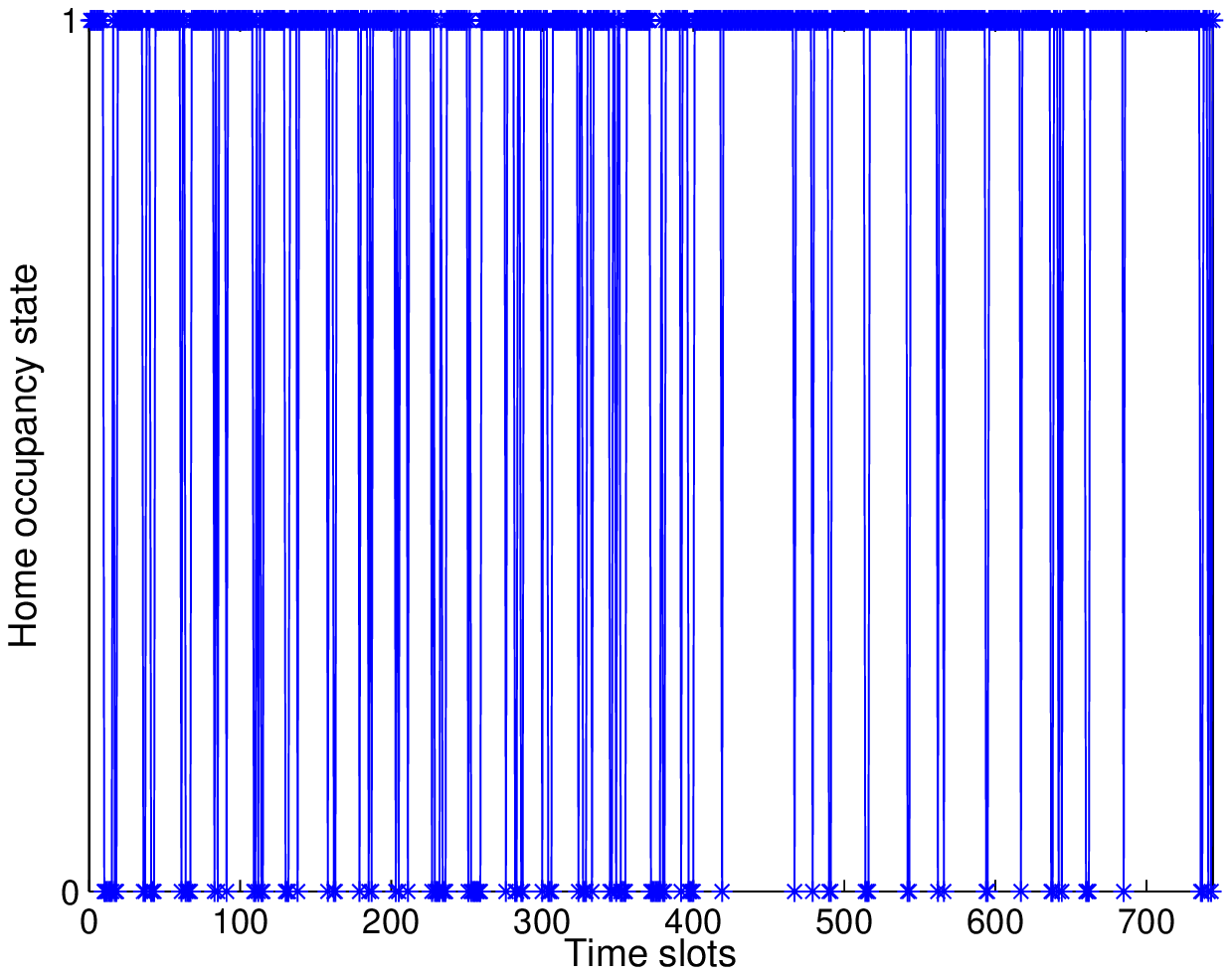}
\end{minipage}
}
\caption{Random input data used in simulations.} \label{fig_2}
\end{figure}

The main simulation parameters are given as follows: $T=744$ hours, $\tau=$1 hour, $T^{\max}=25^oC$, $T^{\text{ref}}_{t+1}=22.5^oC$ for home occupancy, $\eta=1$, $A=\frac{1}{15} kW/^\circ F$\cite{Deng2016}, $S_0=22.5^oC$, $V=\min\{V_1^{\max},V_2^{\max}\}$, $\Gamma=\Gamma^{\max}$ (we set $\Gamma^{\min}=\Gamma^{\max}$), $\alpha=\alpha^{\max}$, $S_t=0.9B_t$\cite{Zhangyu2013}, $\theta_{\text{pv}}=0.2$\cite{Li2016}, $C_{\text{pv}}=30 m^2$, $G^{\max}=20 kWh$\cite{Guo2013}, $G^{\min}=5 kWh$, $u^{\text{dmax}}=u^{\text{cmax}}=1 kW$\cite{Guo2013}, $\varepsilon=0.985$, $v^{\max}=3 kW$, $e^{\max}=8 kW$ (the power could support the heating of a room with 136 $m^2$)\footnote{http://item.gome.com.cn/A0006199221-pop8009870115.html?intcmp=list-9000000700-1\_1\_1}. Suppose $E$ follows a uniform discrete distribution with parameters 4 and 18, EV charging time is [7pm, 6am]. Thus, we have $R=5$. According to the Theorem 2, we set $\xi=(2VB^{\max}+v^{\max})/(R-1)$. In simulations, we adopt the hourly outdoor temperature data\footnote{http://data.cma.cn/} and retail electricity price data associated with the Nanjing City of China in January of 2017, which are shown in Figs.~\ref{fig_2}(a)-(b). For renewable generation information, we use the hourly solar radiation data\footnote{http://midcdmz.nrel.gov/srrl\_bms/historical/} associated with the Golden city of the USA in January of 2017. Due to the lack of home occupancy traces, we adopt the sport data related to the step number in January of 2017 to approximate home occupancy states, which was collected by an iPhone automatically. To be specific, if the total step number during an hour is larger than a given threshold, the home is assumed to be unoccupied. Otherwise, the home is occupied. In this paper, we set the threshold as 1800 (i.e., 2 seconds per step)\footnote{Though the above threshold-based decision may mistake other cases (e.g., office occupancy) for home occupancy, it is still effective for the hours of sleep, e.g., 22pm-7am.} and the obtained home occupancy trace is shown in Fig.~\ref{fig_2}(d).

To show the effectiveness of the proposed algorithm, three baselines are adopted as follows:
\begin{itemize}
\item Baseline-1 (\emph{B1}): similar to \cite{Thatte2012},\cite{Hao2017}, \emph{B1} intends to maintain the most comfortable temperature level $T^{\text{ref}}_{t+1}$ for occupants by drawing the power $e_t=\max(0,\min(e^{\max},\frac{A}{\eta}\Big(\frac{T^{\text{ref}}_{t+1}-\varepsilon T_t}{1-\varepsilon}-T_t^{\text{out}}\Big)))$ when the home is occupied. When $\pi_{t+1}=0$ and $T_{t+1}|_{e_{t}=0}\geq T^{\min}$, we set $e_t=0$. In addition, \emph{B1} does not consider ESS and serves EV charging demand instantly.
\item Baseline-2 (\emph{B2}): \emph{B2} intends to implement the temporally-decoupled algorithm as in \cite{Deng2016} with perfect one-step price forecasting when the home is occupied, which results in an optimal HVAC control if \eqref{f_2} is neglected. Moreover, random occupancy is also considered in \emph{B2}, i.e., when $\pi_{t+1}=0$ and $T_{t+1}|_{e_{t}=0}\geq T^{\min}$, we set $e_t=0$. In addition, \emph{B2} does not consider ESS and serves EV charging demand instantly.
\item Baseline-3 (\emph{B3}): the proposed algorithm is equivalent to \emph{B3} when ESS control is not considered (i.e., $u^{\text{dmax}}=u^{\text{cmax}}=0 kW$).
\end{itemize}

\begin{figure}
\centering
\subfigure[Indoor temperature (no vacancy)]{
\begin{minipage}[b]{0.4\textwidth}
\includegraphics[width=1\textwidth]{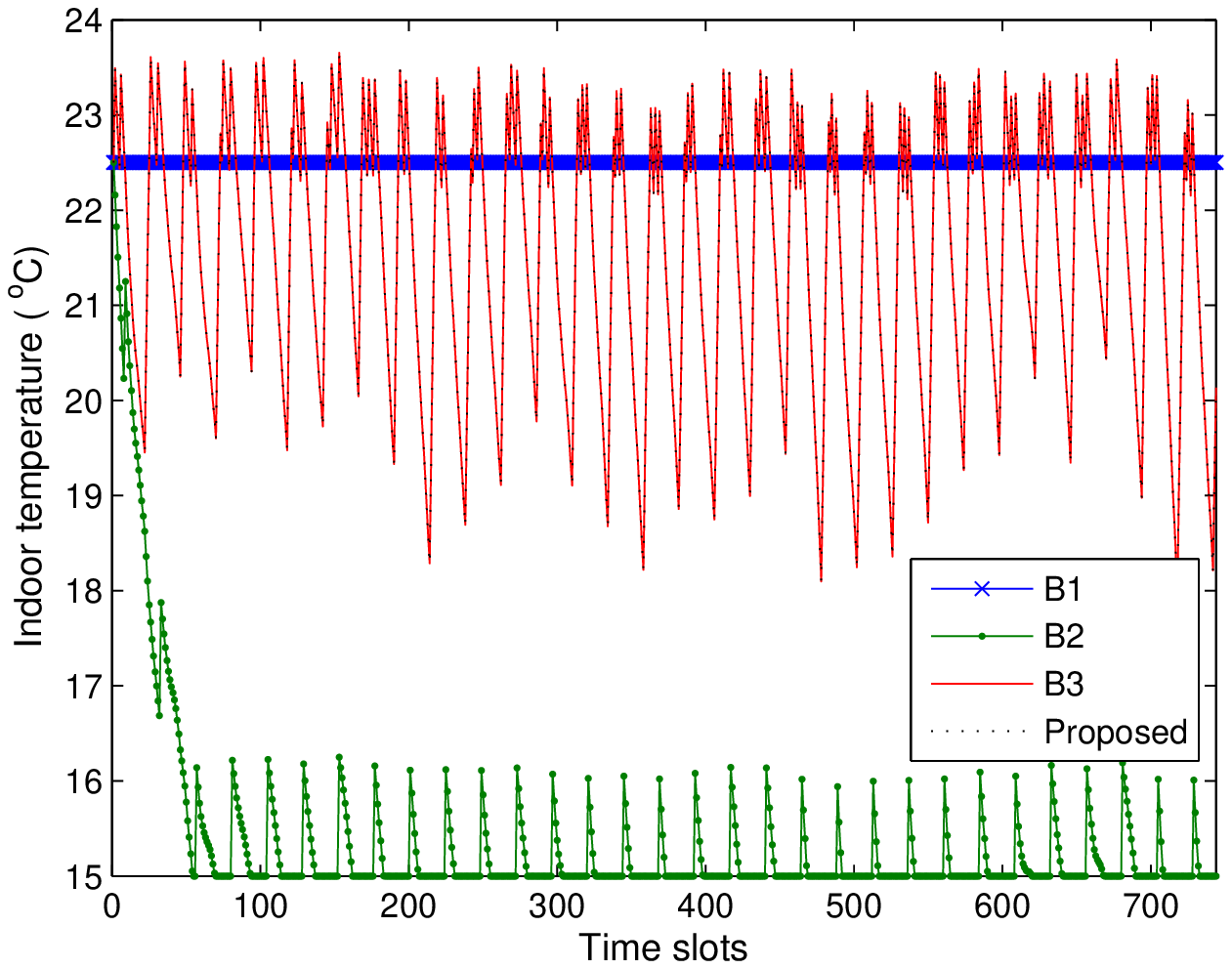}
\end{minipage}
}\\
\subfigure[Indoor temperature (with vacancy)]{
\begin{minipage}[b]{0.4\textwidth}
\includegraphics[width=1\textwidth]{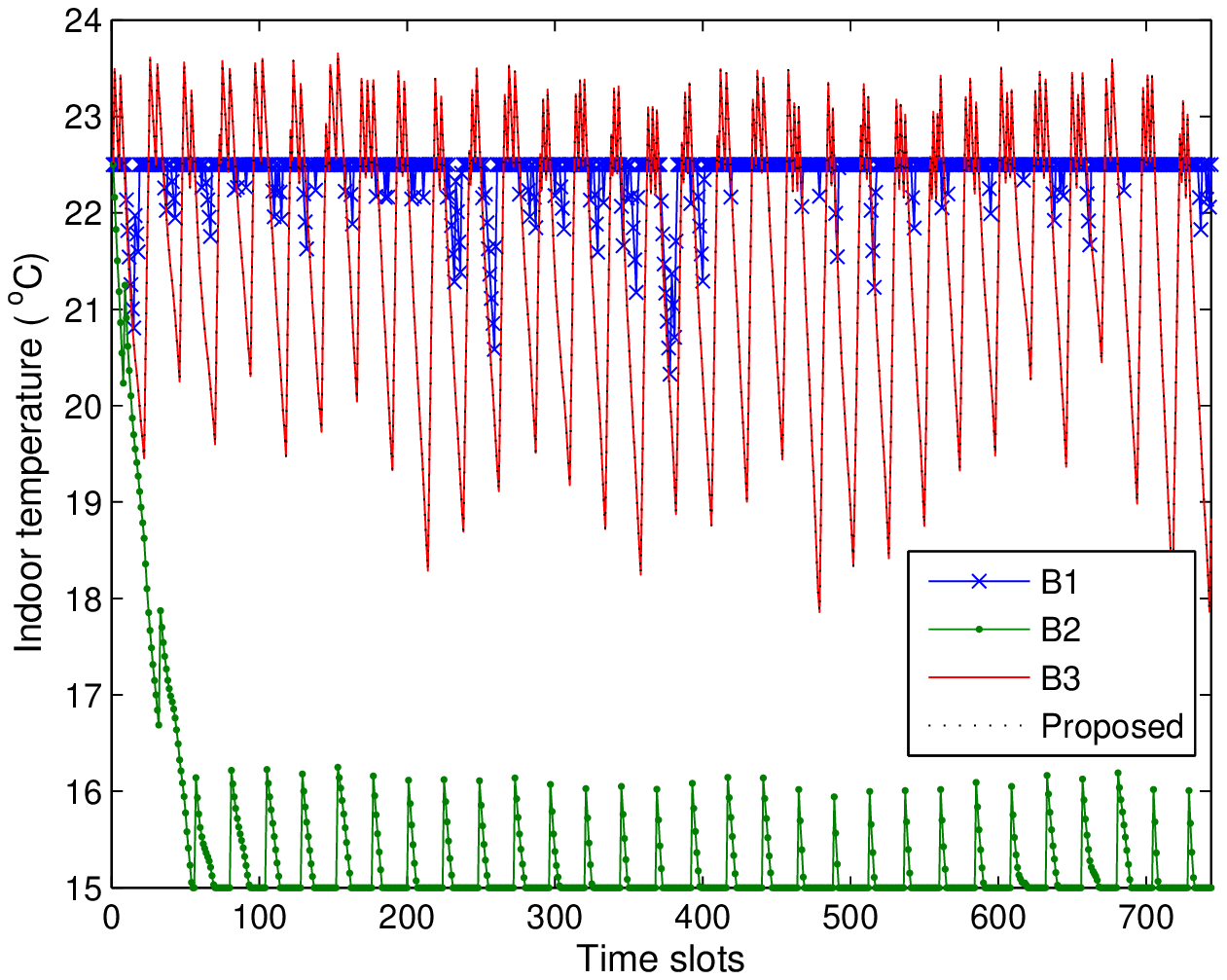}
\end{minipage}
}\\
\subfigure[ESS energy level]{
\begin{minipage}[b]{0.4\textwidth}
\includegraphics[width=1\textwidth]{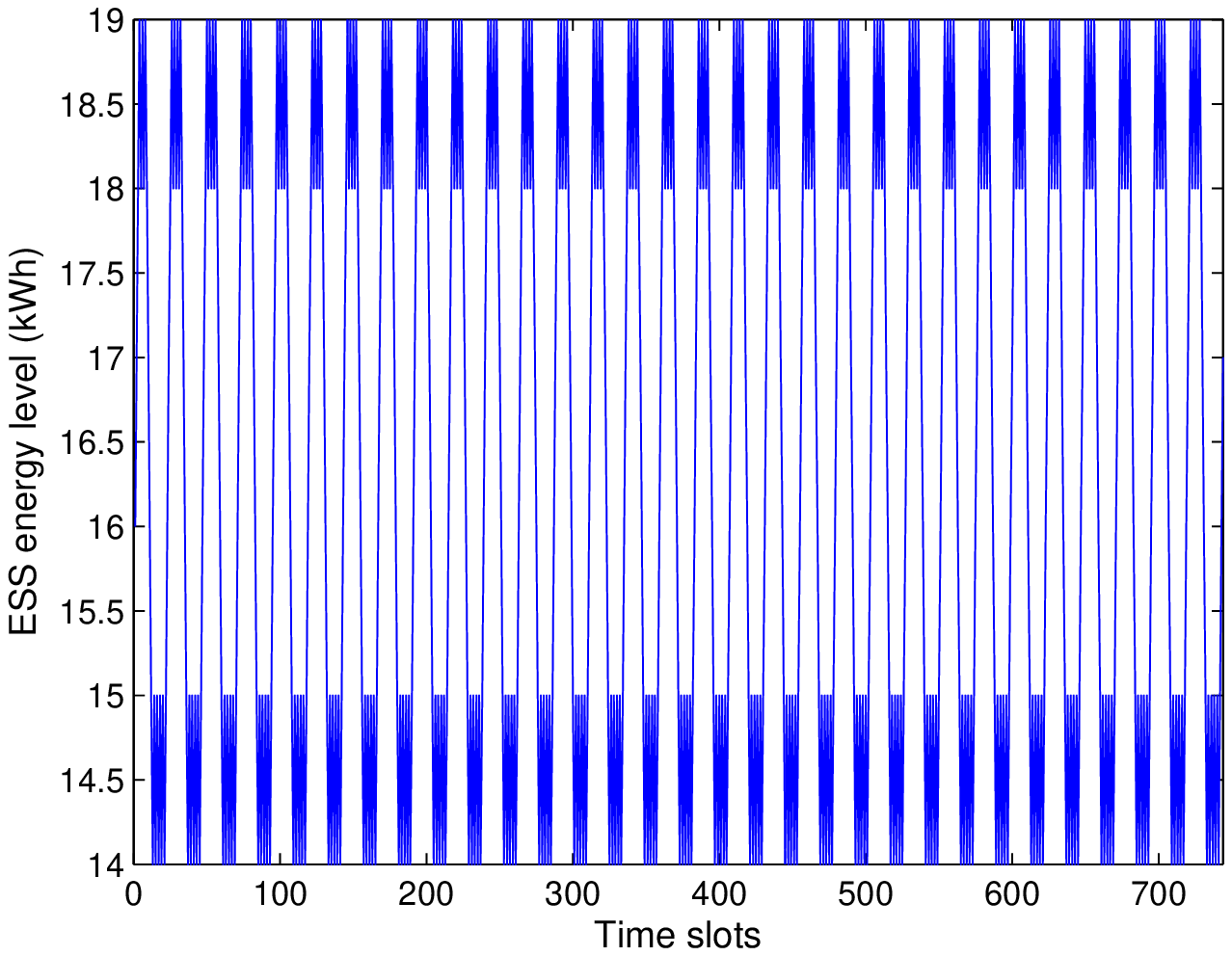}
\end{minipage}
}\\
\subfigure[EV charging delay]{
\begin{minipage}[b]{0.4\textwidth}
\includegraphics[width=1\textwidth]{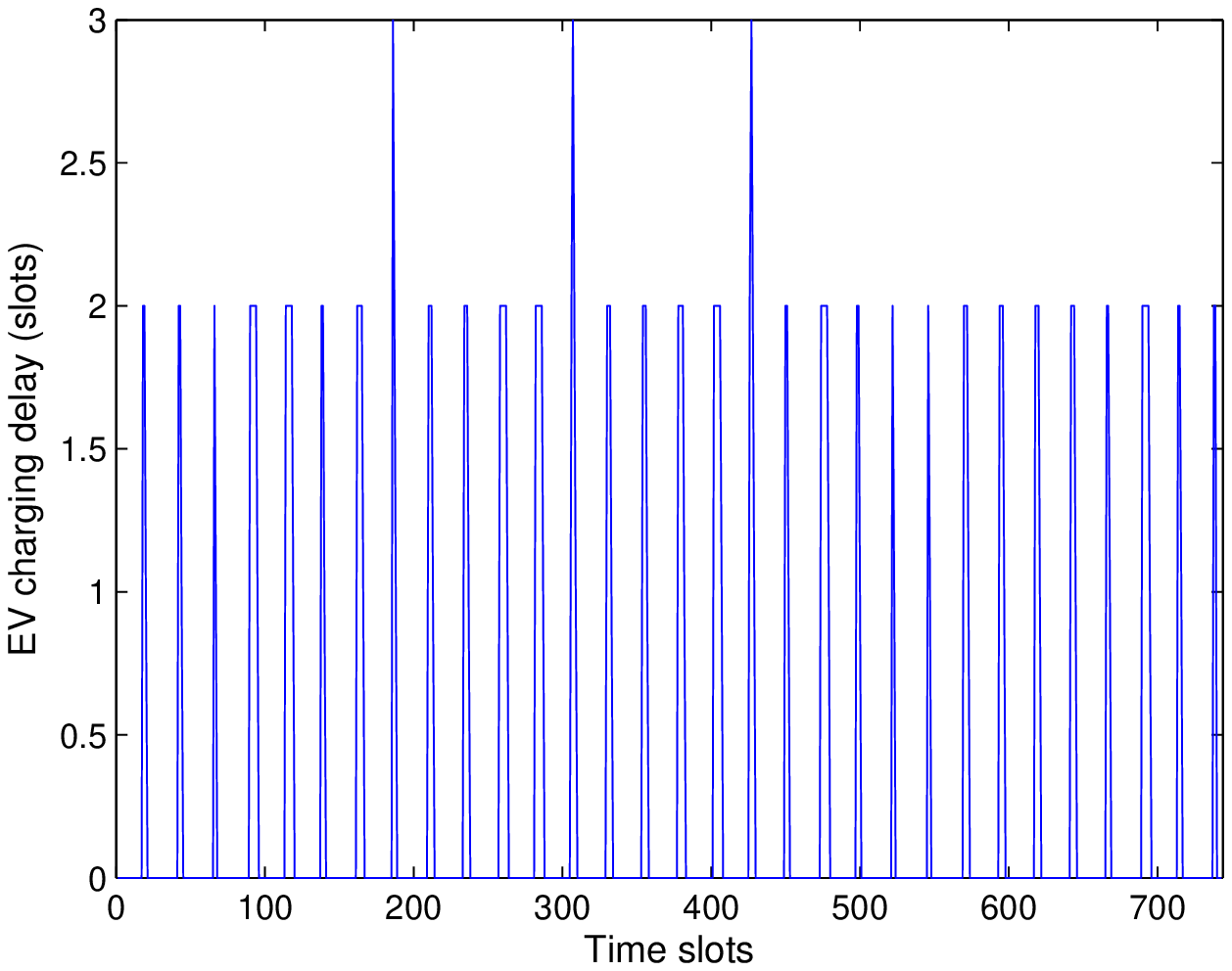}
\end{minipage}
}
\caption{The feasibility of the proposed algorithm (given $\varepsilon=0.98$, $T^{\min}=15^oC$, $\gamma=0$).} \label{fig_3}
\end{figure}

\begin{figure}
\centering
\subfigure[Energy cost]{
\begin{minipage}[b]{0.4\textwidth}
\includegraphics[width=1\textwidth]{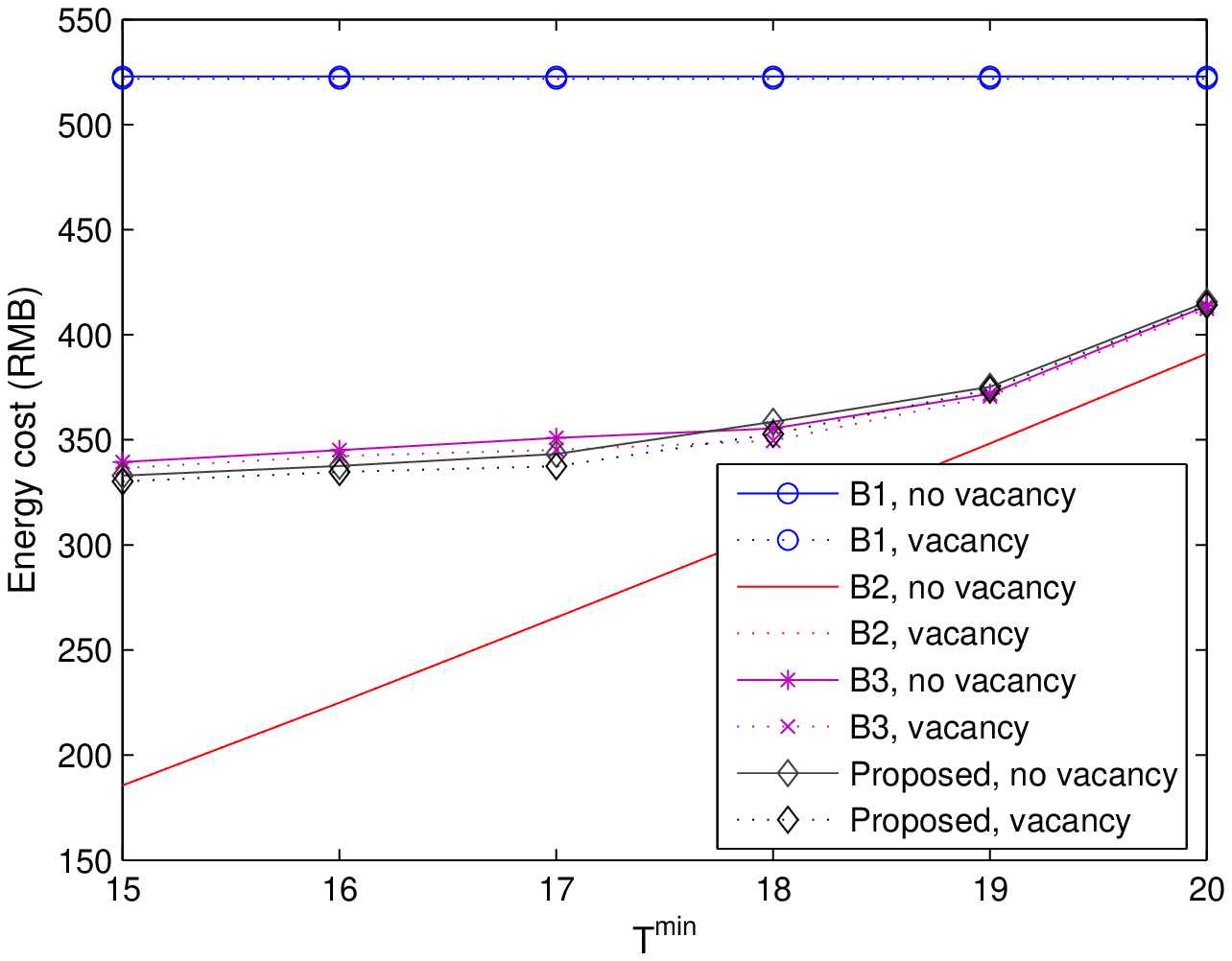}
\end{minipage}
}\\
\subfigure[ATD]{
\begin{minipage}[b]{0.4\textwidth}
\includegraphics[width=1\textwidth]{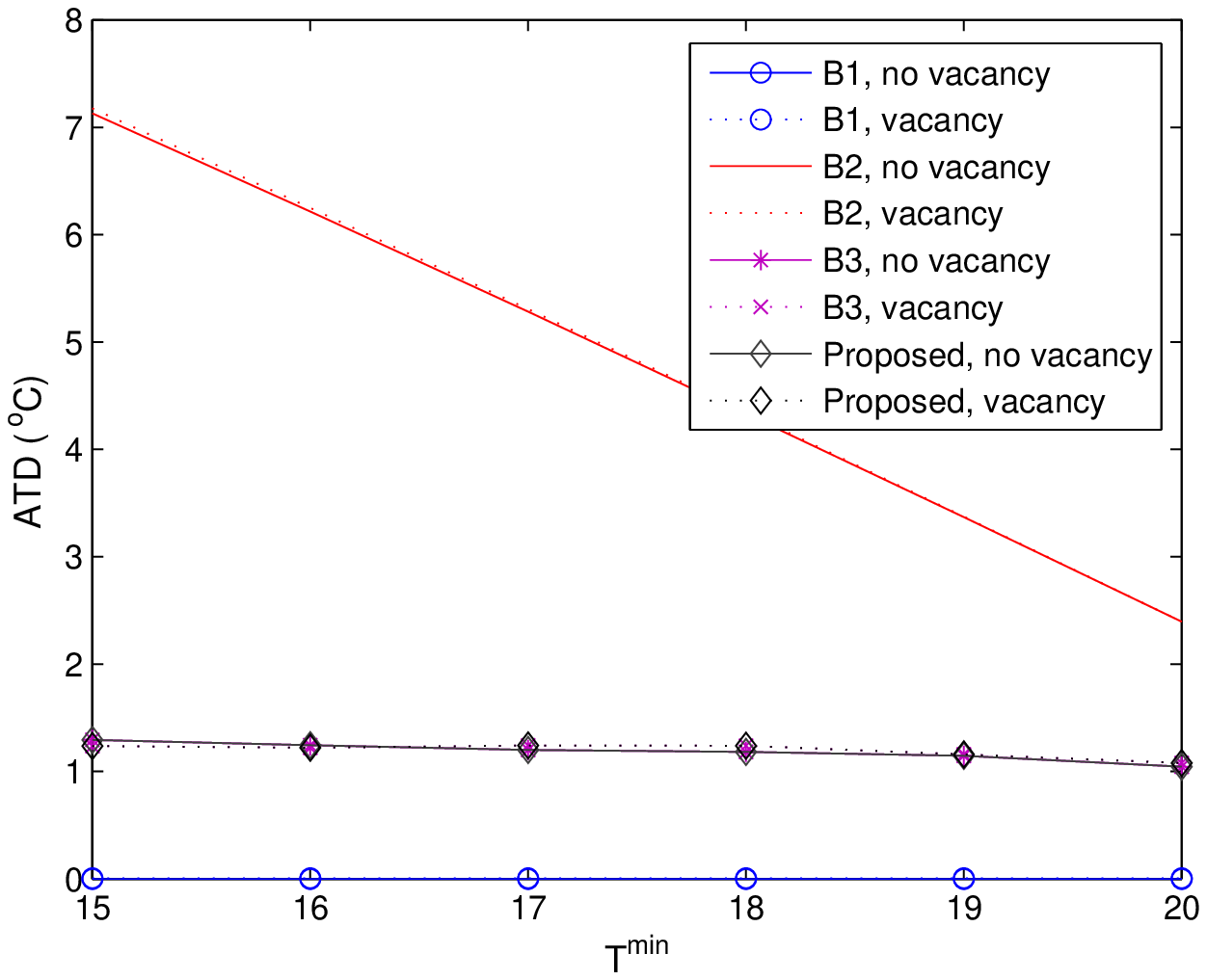}
\end{minipage}
}\\
\subfigure[Actual temperature range]{
\begin{minipage}[b]{0.4\textwidth}
\includegraphics[width=1\textwidth]{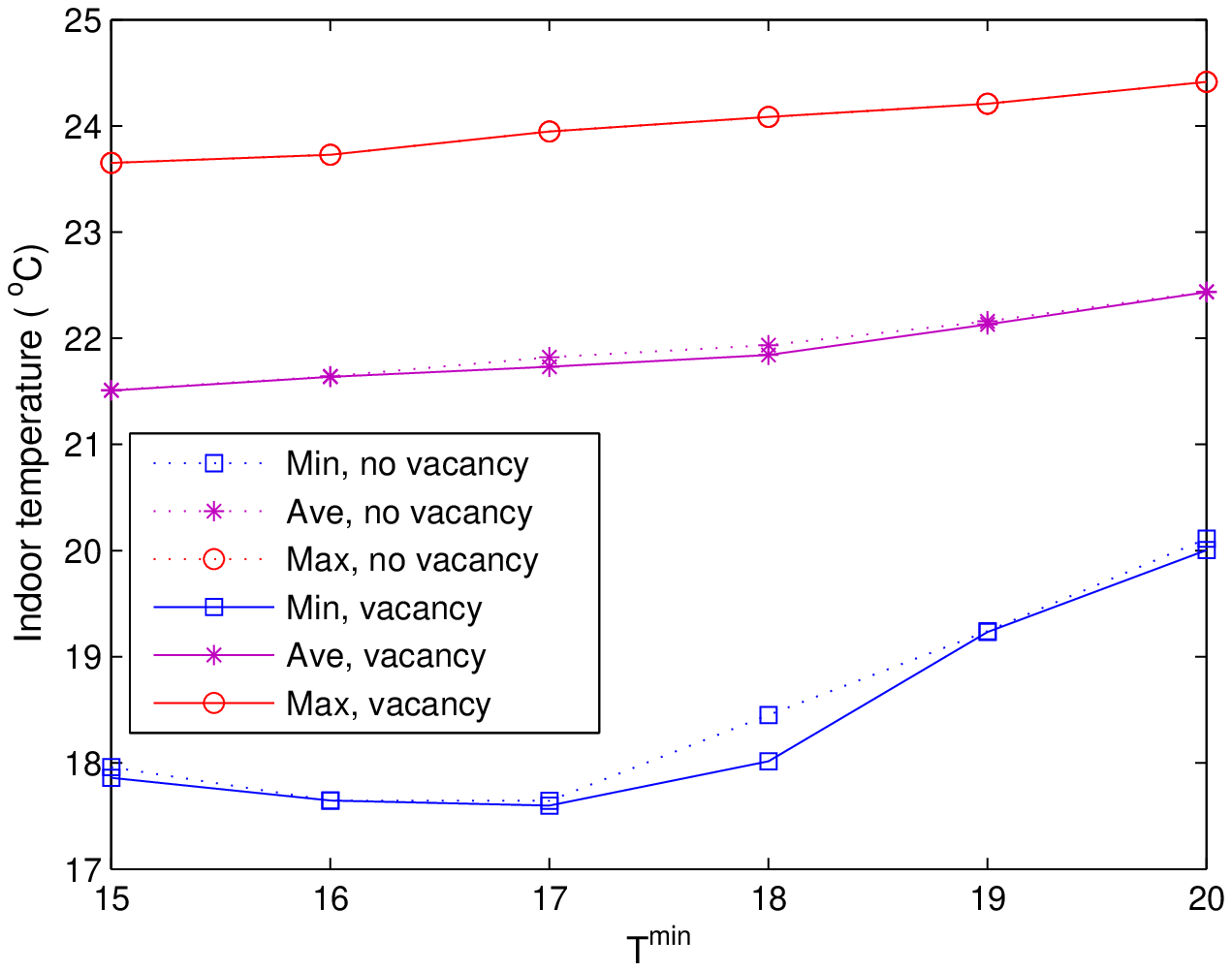}
\end{minipage}
}\\
\subfigure[$V=\min\{V_1^{\max},V_2^{\max}\}$]{
\begin{minipage}[b]{0.4\textwidth}
\includegraphics[width=1\textwidth]{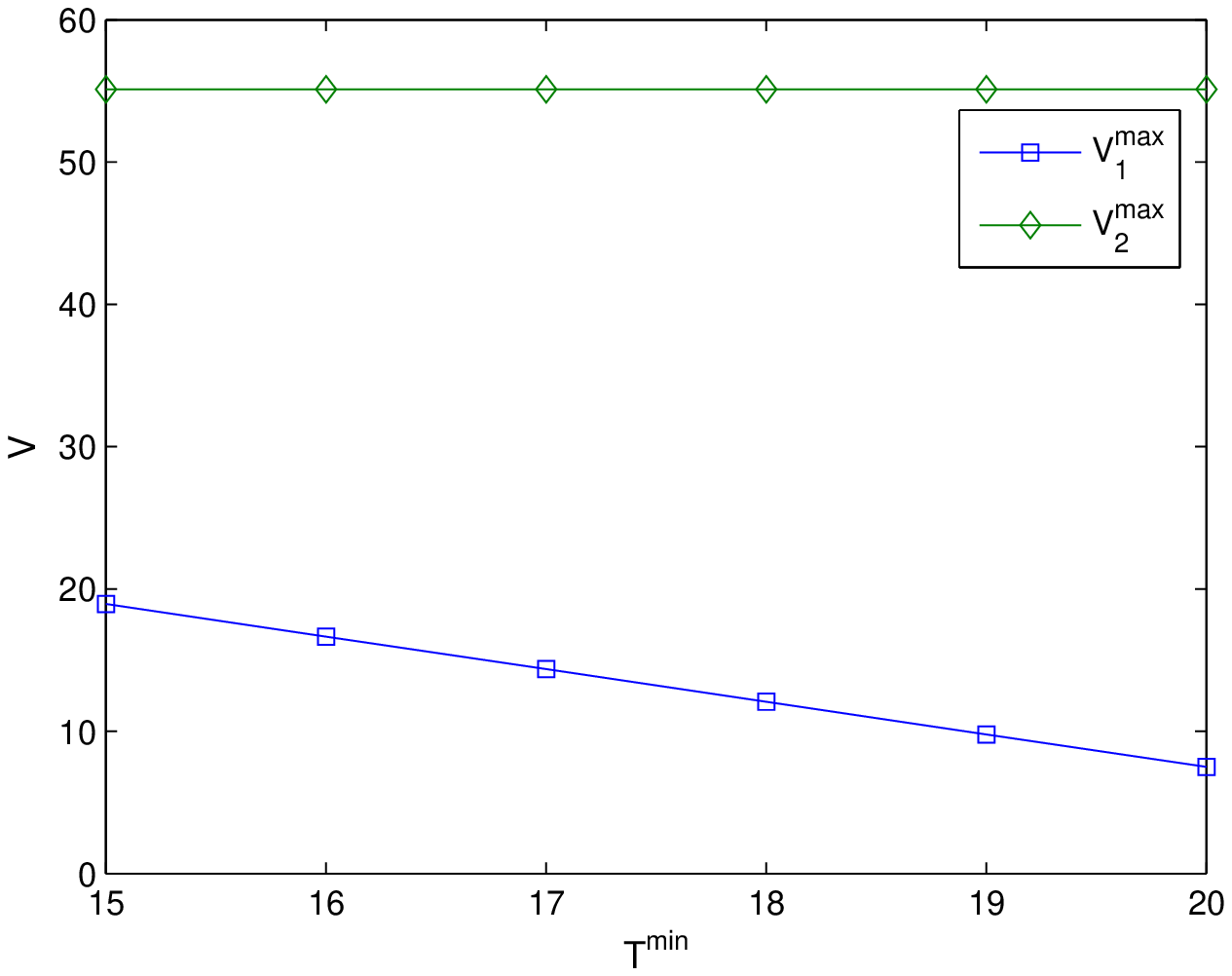}
\end{minipage}
}
\caption{The impact of $T^{\min}$ (given $\varepsilon=0.98$,~$\gamma=0$).} \label{fig_4}
\end{figure}

\begin{figure}
\centering
\subfigure[Energy cost]{
\begin{minipage}[b]{0.4\textwidth}
\includegraphics[width=1\textwidth]{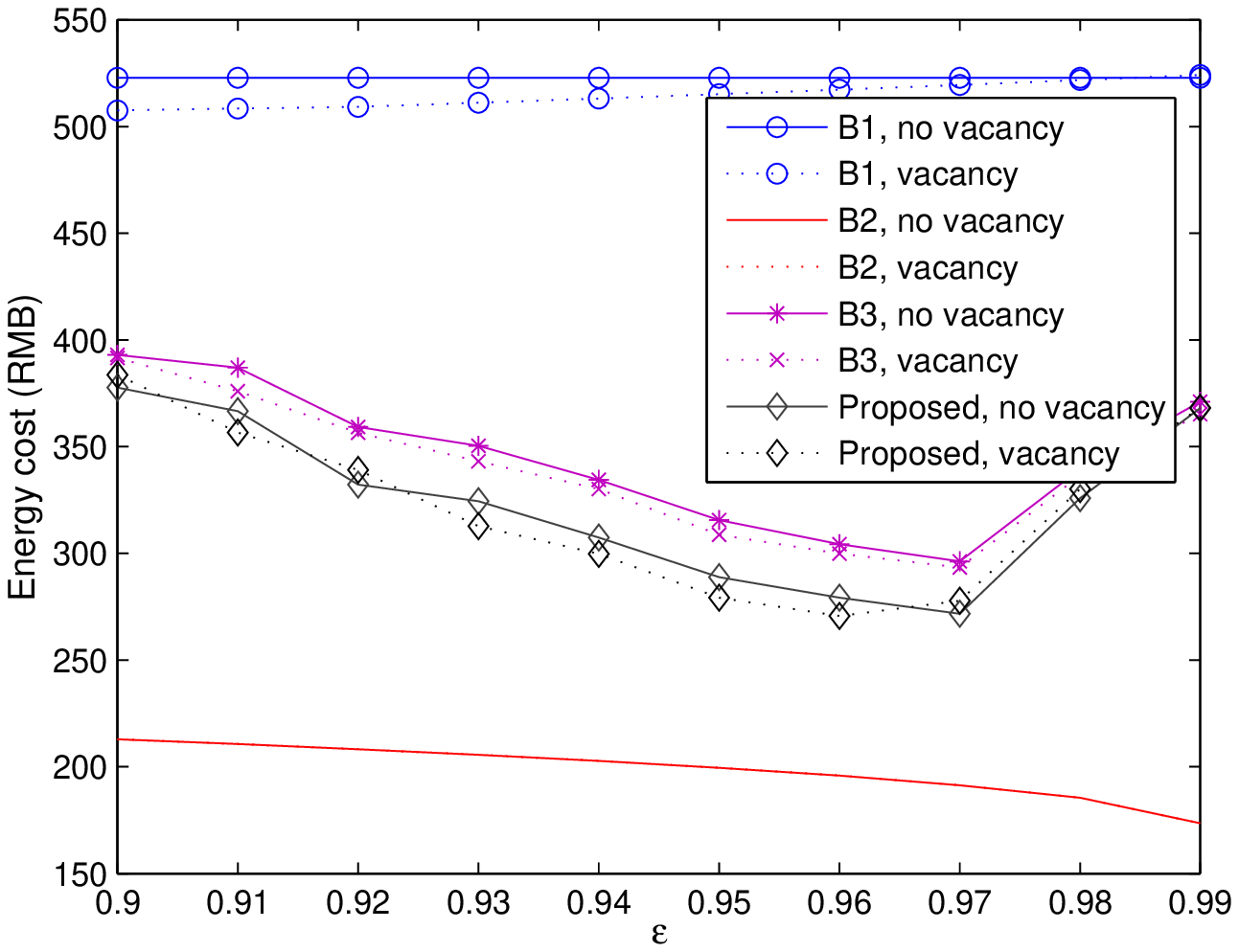}
\end{minipage}
}\\
\subfigure[ATD]{
\begin{minipage}[b]{0.4\textwidth}
\includegraphics[width=1\textwidth]{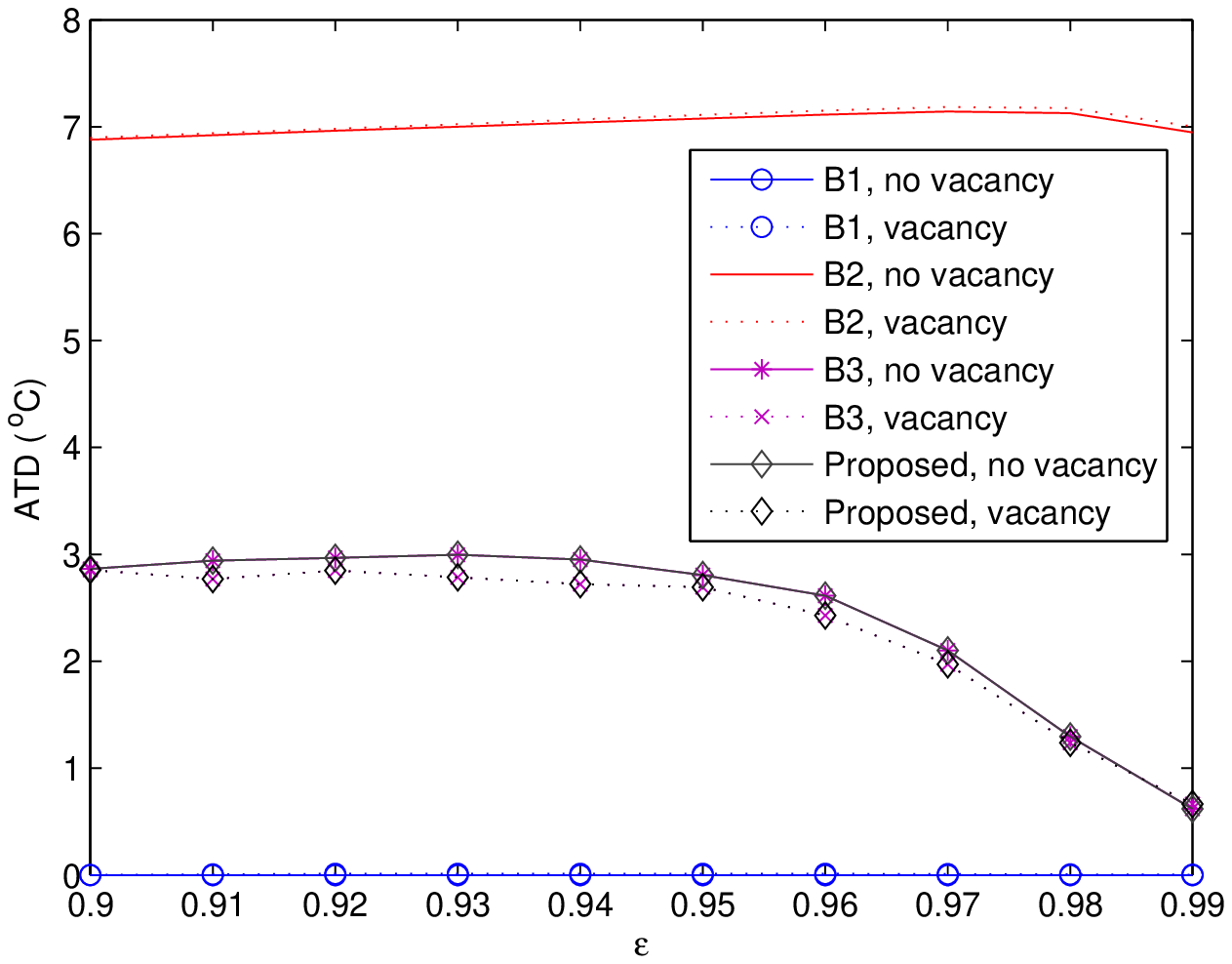}
\end{minipage}
}\\
\subfigure[Actual temperature range]{
\begin{minipage}[b]{0.4\textwidth}
\includegraphics[width=1\textwidth]{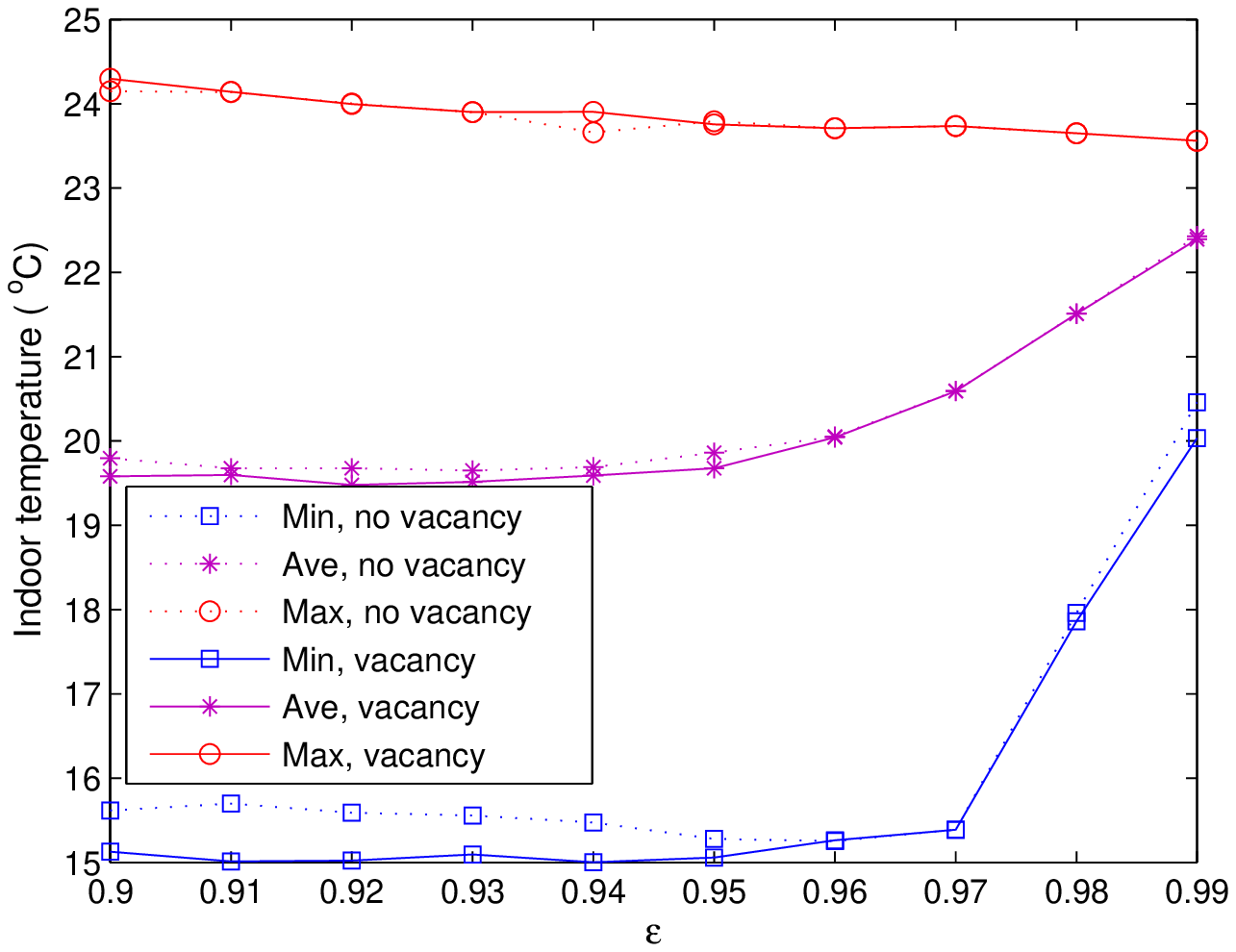}
\end{minipage}
}\\
\subfigure[$V=\min\{V_1^{\max},V_2^{\max}\}$]{
\begin{minipage}[b]{0.4\textwidth}
\includegraphics[width=1\textwidth]{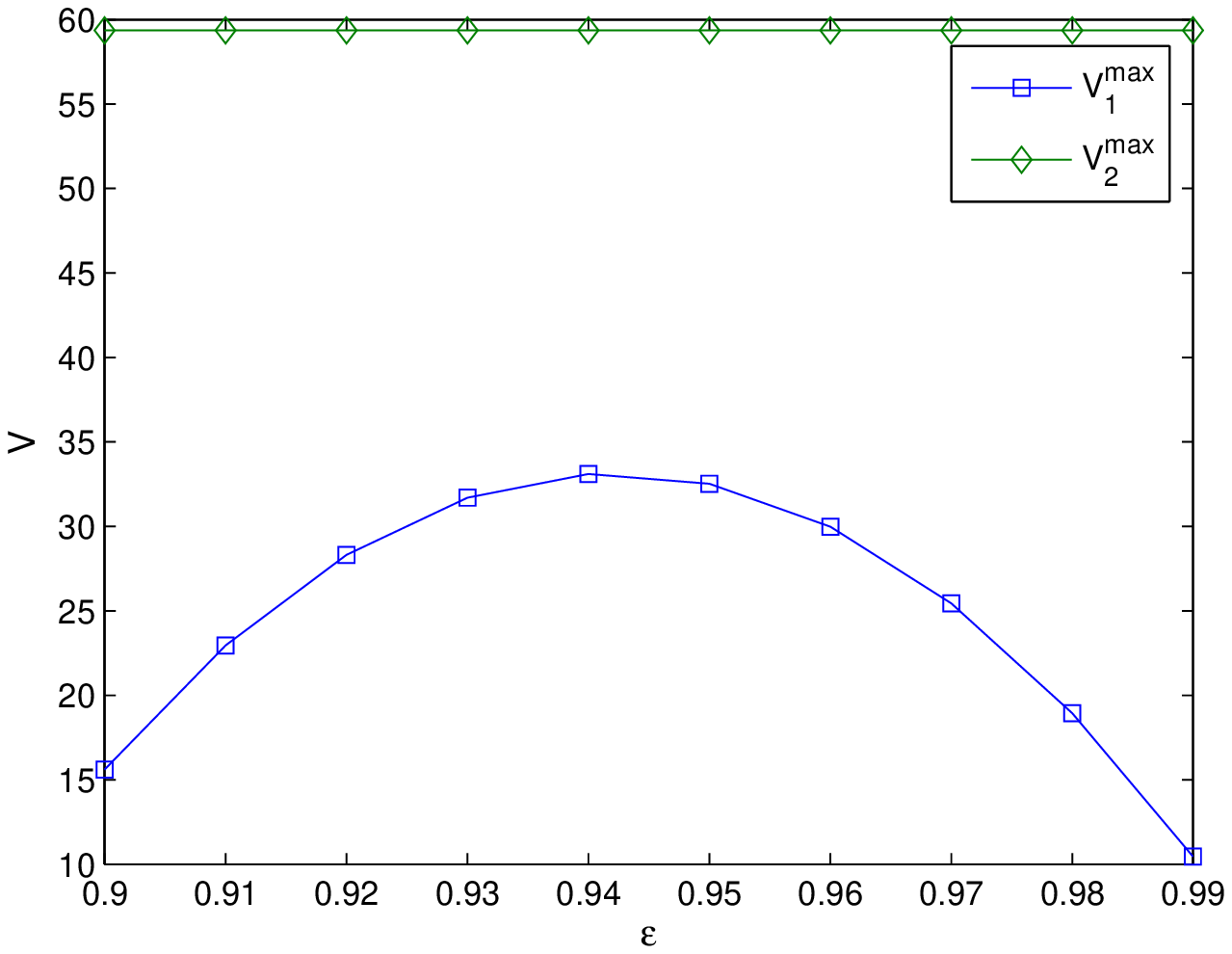}
\end{minipage}
}
\caption{The impact of $\varepsilon$ (given $T^{\min}=15^oC$, $\gamma=0$).} \label{fig_5}
\end{figure}

\begin{figure}
\centering
\subfigure[Total cost]{
\begin{minipage}[b]{0.4\textwidth}
\includegraphics[width=1\textwidth]{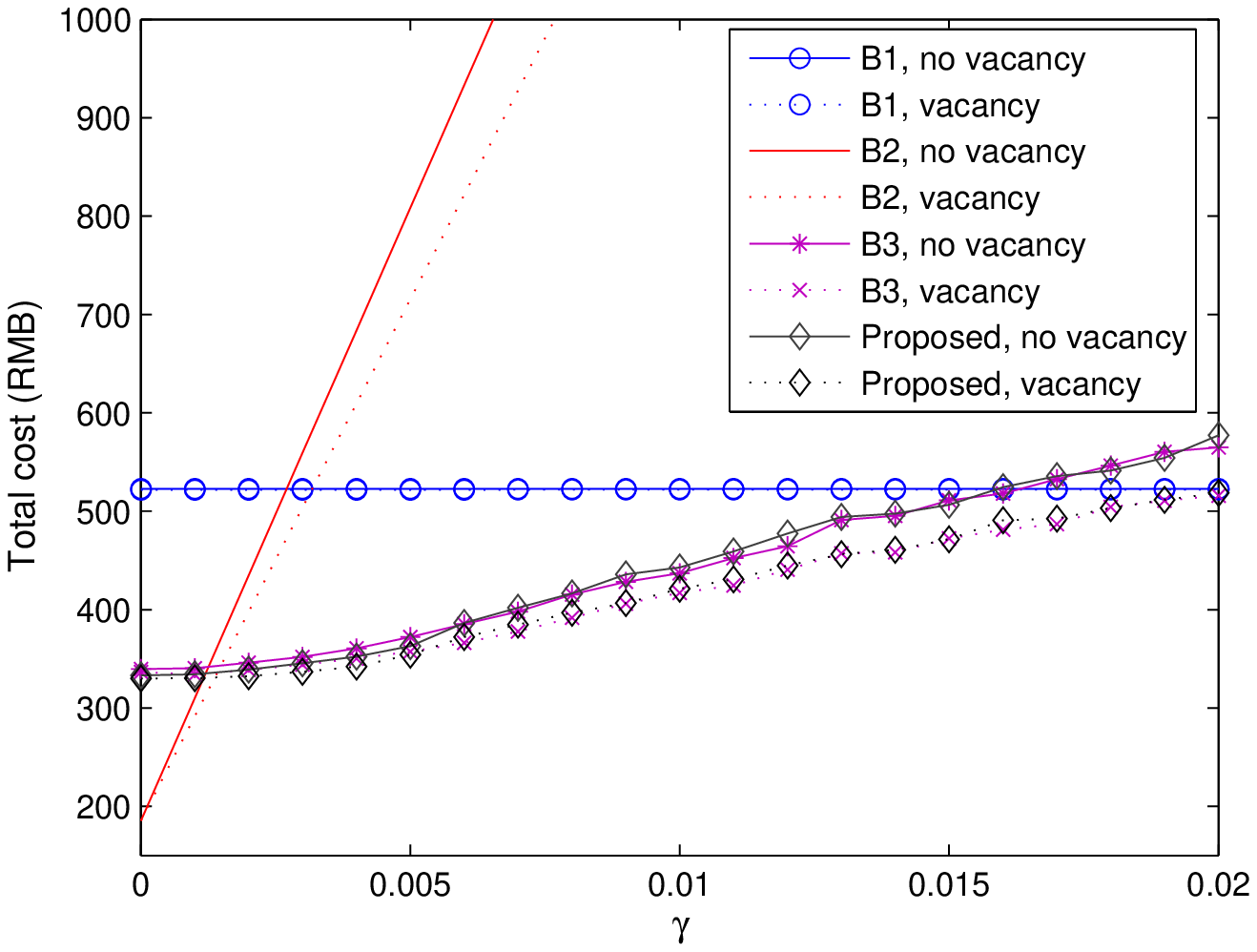}
\end{minipage}
}\\
\subfigure[Energy cost]{
\begin{minipage}[b]{0.4\textwidth}
\includegraphics[width=1\textwidth]{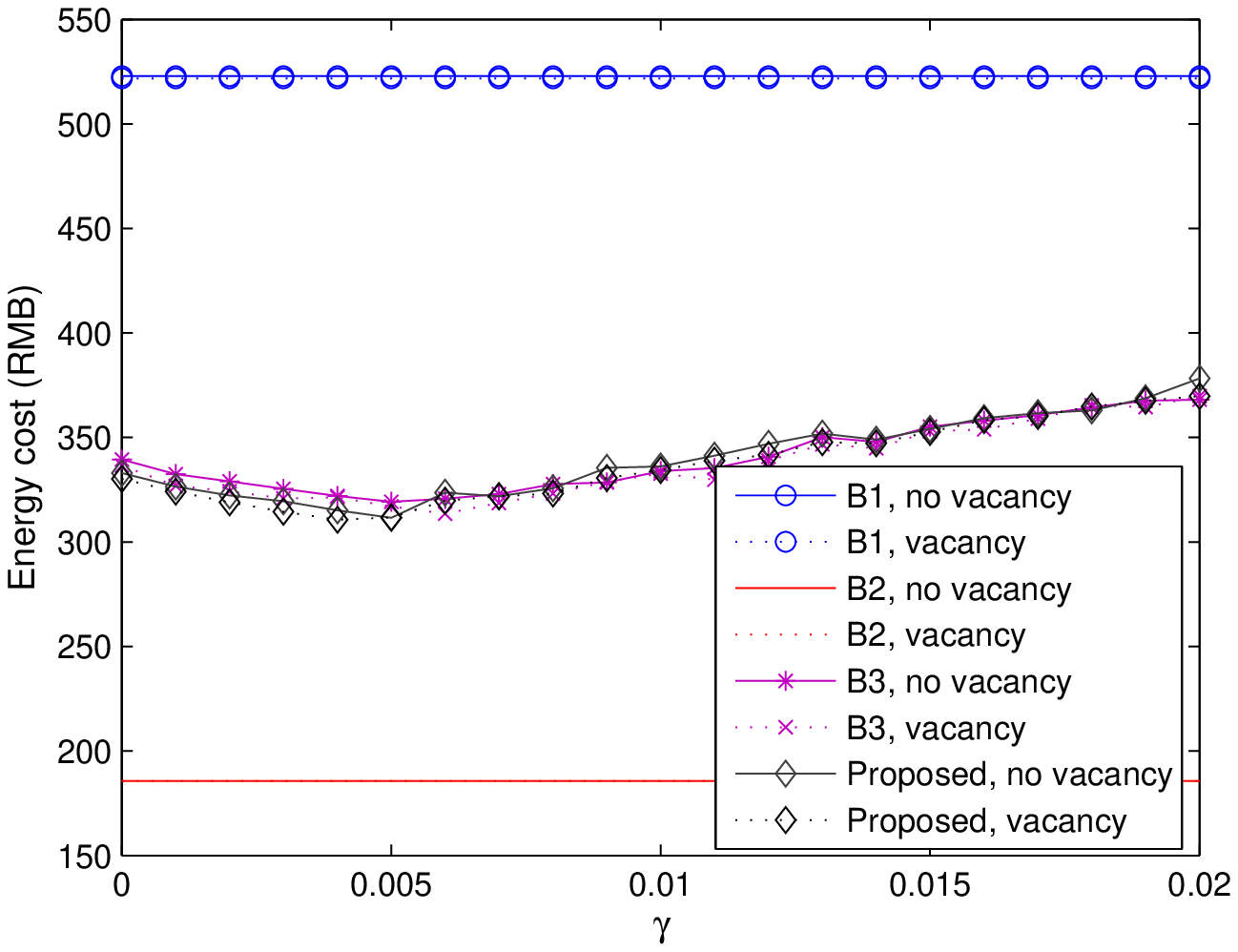}
\end{minipage}
}\\
\subfigure[Thermal discomfort cost]{
\begin{minipage}[b]{0.4\textwidth}
\includegraphics[width=1\textwidth]{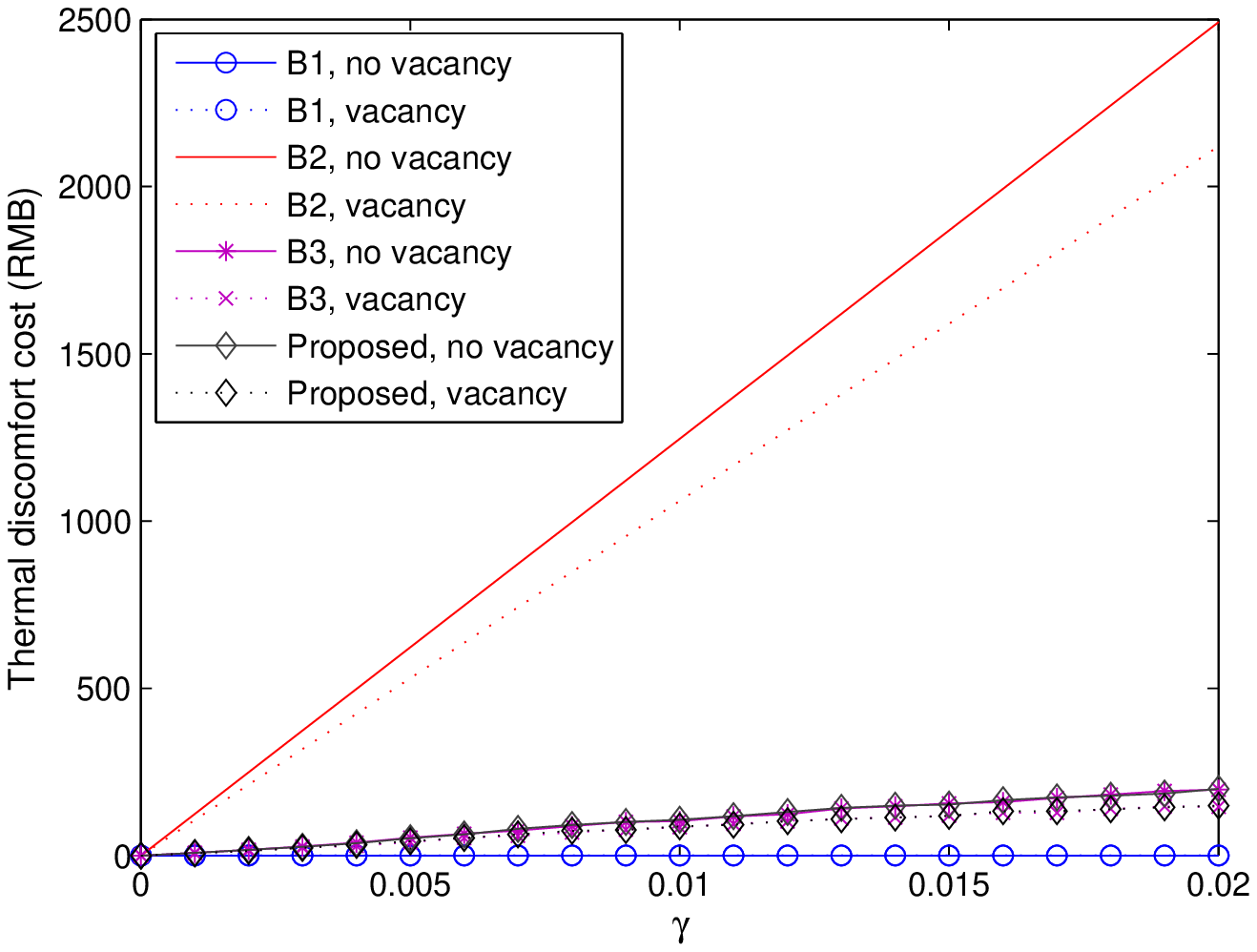}
\end{minipage}
}\\
\subfigure[ATD]{
\begin{minipage}[b]{0.4\textwidth}
\includegraphics[width=1\textwidth]{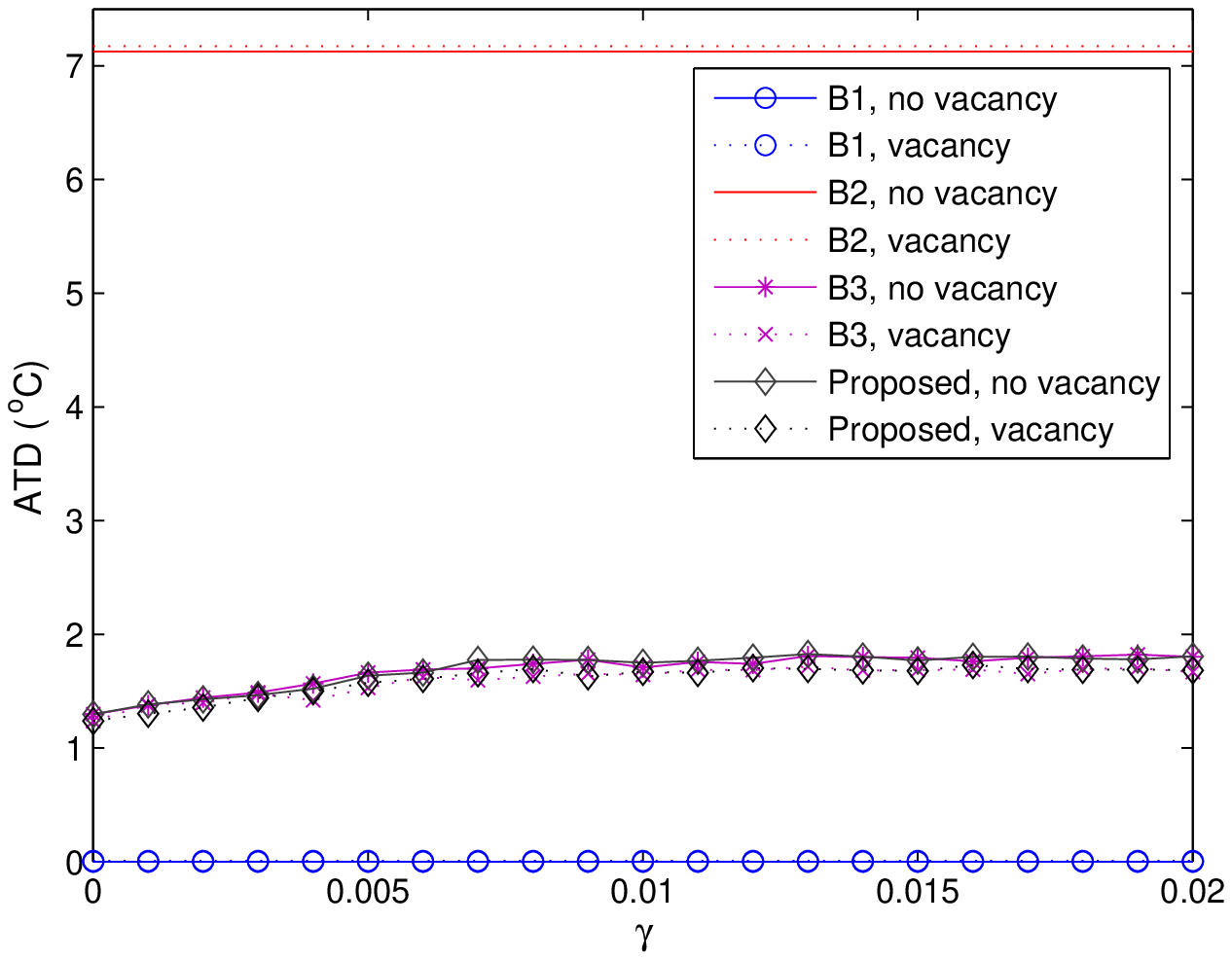}
\end{minipage}
}
\caption{The impact of $\gamma$ (given $\varepsilon=0.98$,~$T^{\min}=15^oC$).} \label{fig_6}
\end{figure}

\subsection{Simulation results}
\subsubsection{Algorithm feasibility}
According to Theorems 1-3, the algorithmic feasibility of the proposed algorithm could be verified by checking the normal ranges of indoor temperature, ESS energy level, and EV charging delay. As shown in Fig.~\ref{fig_3}, indoor temperature under the proposed algorithm and ESS energy level always fluctuate within their respective normal ranges. Moreover, the maximum EV charging delay is less than $R=5$. Therefore, the proposed algorithm is feasible to the original problem \textbf{P1}. In addition, it can be observed that B1 intends to maintain the most comfortable level when the home is occupied. When the home is unoccupied, B1 will turn off the HVAC system if that decision would not result in a temperature below $T^{\min}$ in the next time slot. For B2, the increase or decrease of the indoor temperature depends on the relationship between $B_t$ and $B_{t+1}$. When $B_t<\varepsilon B_{t+1}$, the temperature will increase for pre-heating. Otherwise, the indoor temperature will decrease.

\subsubsection{The impact of $T^{\min}$}
Since the proposed algorithm adjusts the HVAC power input dynamically according to the current electricity price, larger temperature range would result in lower energy cost. With the increase of $T^{\min}$, normal temperature range ($T^{\max}-T^{\min}$) decreases. Consequently, the proposed algorithm and B3 achieve higher energy cost but lower ATD (i.e., Average Temperature Deviation from the most comfortable level $T^{\text{ref}}_{t+1}$: $\text{ATD}=\frac{1}{N_{\text{on}}-1}\sum\nolimits_{t=0}^{N-2}|T_{t+1}-T^{\text{ref}}_{t+1}|\pi_{t+1}$, and $N_{\text{on}}$ denotes the total number of time slots with home occupancy) as shown in Fig.~\ref{fig_4}. Moreover, the proposed algorithm could reduce energy cost by 36.3\% compared with B1 with small sacrifice of ATD (ATD=1.295). When $V\geq 13$, the proposed algorithm achieves lower energy cost than B3. Otherwise, B3 achieves lower energy cost than the proposed algorithm. The reason is that a small control parameter $V$ would affect the utilization of temporal price diversity as shown in \cite{LiangYuPowerOutage2015}.

\subsubsection{The impact of $\varepsilon$}
The performance of the proposed algorithm under varying $\varepsilon$ is demonstrated by Fig.~\ref{fig_5}, where the proposed algorithm generally achieves lower energy cost given a larger $\varepsilon$. The reason is that larger $\varepsilon$ would result in less thermal loss given the same time horizon, which contributes to reaping the benefits of temporal price diversity under our proposed algorithm. Here, the temporal price diversity means that the proposed algorithm would increase power inputs when electricity prices are low so that power inputs associated with high electricity prices in later time slots could be reduced, resulting in lower energy cost. When $\varepsilon\geq 0.98$, the energy cost begins to increase. The reason is that a small $V$ will lead to a small actual temperature range as shown in Fig.~\ref{fig_5}(c). Compared with B3, the proposed algorithm could reduce energy cost without the sacrifice of ATD under the given configuration. Though B2 has the lowest energy cost, its ATD is also the largest. Therefore, B2 is not necessarily the best scheme when energy cost and thermal discomfort cost are jointly considered.

\subsubsection{The impact of $\gamma$}
Fig.~\ref{fig_6} illustrates the performance of the proposed algorithm and three baselines under varying $\gamma$. It can be found that the proposed algorithm achieves the lowest total cost when $\gamma$ falls within an appropriate range, e.g., $[0.002,~0.016]$ (no vacancy) and $[0.002,~0.02]$ (with vacancy). Specifically, when the tolerant ATD is smaller than $2^oC$, the proposed algorithm could achieve lower energy cost than \emph{B1} by 40.41\% and 38.95\% when the home is always occupied and randomly occupied, respectively. In contrast, \emph{B2} achieves the best performance in a much smaller range $[0,~0.001]$ since its thermal discomfort cost would be the largest given a larger $\gamma$ as shown in Fig.~\ref{fig_6}(c). When $\gamma$ is large enough (e.g., $\gamma>0.02$), \emph{B1} would achieve the best performance since the thermal discomfort cost of \emph{B1} is the smallest one and the corresponding energy cost is a constant, which are illustrated by Figs.~\ref{fig_6}(b) and (c). In summary, the proposed algorithm offers an effective way of controlling the HVAC system when home occupants care about both energy cost and thermal comfort.

\section{Conclusions and Future Work}\label{s6}
In this paper, we investigated the energy management of a sustainable smart home with an HVAC load and random occupancy. To minimize the sum of energy cost and thermal discomfort cost in a long-term time horizon, we proposed an online energy management algorithm based on the LOT framework without predicting any system parameters. Different from existing Lyapunov-based energy management algorithms, the proposed algorithm does not require submitting unknown power demands of an HVAC system into an energy queue. Extensive simulation results based on real-world traces showed the effectiveness of the proposed algorithm. In the future, we plan to investigate the online HVAC control in a commercial building\cite{Hao2017}, e.g., how to allocate the air supply rate of every zone or room in realtime while taking thermal discomfort of occupants into consideration. Moreover, we plan to investigate the impact of HVAC load aggregation\cite{Erdi2017} in a residential building on end-user comfort, e.g., given some households enrolled in a demand response program, how to minimize the average thermal discomfort of these households without violating the total power reduction/increase requirement during a demand response event.

\appendices

\section{Proof of Lemma 1}
\begin{IEEEproof}
Let ($e_t^*$,$x_t^*$,$y_t^*$,$g_t^*$) be the optimal solution of \textbf{P2}.
\begin{enumerate}
  \item When $VS^{\min}+b_t>-\varepsilon(1-\varepsilon)H_t\frac{\eta}{A}$, suppose $e_t^*>0$. To prove that the above assumption does not hold, we construct another solution (0,$x_t^*$,$y_t^*$,$g_t^\diamond$). According to power balance, we have $g_t^\diamond=g_t^*-e_t^*$. Let the optimal objective value of \textbf{P2} be $\Omega_2$ and the objective value with the new solution (0,$x_t^*$,$y_t^*$,$g_t^\diamond$) be $\Omega_2^\diamond$, respectively.
      Then, we can compare $\Omega_2$ with $\Omega_2^\diamond$ under different symbols of $g_t^\diamond$ and $g_t^*$ as follows.
      \begin{itemize}
        \item When $g_t^*<0$, we have $g_t^\diamond <0$ since $g_t^\diamond=g_t^*-e_t^*$ and $e_t^*>0$. Then, $\Omega_2-\Omega_2^\diamond=(VS_t+\varepsilon(1-\varepsilon)H_t\frac{\eta}{A}+b_t)e_t^*>(VS^{\min}+\varepsilon(1-\varepsilon)H_t\frac{\eta}{A}+b_t)e_t^*>0$.
        \item When $g_t^*>0$ and $g_t^\diamond>0$. Then, $\Omega_2-\Omega_2^\diamond=(VB_t+\varepsilon(1-\varepsilon)H_t\frac{\eta}{A}+b_t)e_t^*>(VS^{\min}+\varepsilon(1-\varepsilon)H_t\frac{\eta}{A}+b_t)e_t^*>0$.
        \item When $g_t^*>0$ and $g_t^\diamond<0$. Then, $\Omega_2-\Omega_2^\diamond>(VB_t+\varepsilon(1-\varepsilon)H_t\frac{\eta}{A}+b_t)e_t^*>(VS^{\min}+\varepsilon(1-\varepsilon)H_t\frac{\eta}{A}+b_t)e_t^*>0$.
      \end{itemize}
      Taking the above three cases into consideration, we have the conclusion that $e_t^*=0$ when $VS^{\min}+b_t>-\varepsilon(1-\varepsilon)H_t\frac{\eta}{A}$.
  \item When $VB^{\max}+c_t<-\varepsilon(1-\varepsilon)H_t\frac{\eta}{A}$, suppose $e_t^*<e^{\max}$. To prove that the above assumption does not hold, we construct another solution ($e^{\max}$,$x_t^*$,$y_t^*$,$g_t^\diamond$). According to power balance, we have $g_t^*-g_t^\diamond=e_t^*-e^{\max}$. Let the optimal objective value of \textbf{P2} be $\Omega_2$ and the objective value with the new solution ($e^{\max}$,$x_t^*$,$y_t^*$,$g_t^\diamond$) be $\Omega_2^\diamond$, respectively.
      Then, we can compare $\Omega_2$ with $\Omega_2^\diamond$ under different symbols of $g_t^\diamond$ and $g_t^*$ as follows.
      \begin{itemize}
        \item When $g_t^*>0$, we have $g_t^\diamond >0$ since $g_t^*<g_t^\diamond$. Then, $\Omega_2-\Omega_2^\diamond>(VB_t+\varepsilon(1-\varepsilon)H_t\frac{\eta}{A}+c_t)(e_t^*-e^{\max})>(VB^{\max}+\varepsilon(1-\varepsilon)H_t\frac{\eta}{A}+c_t)(e_t^*-e^{\max})>0$.
        \item When $g_t^*<0$ and $g_t^\diamond<0$. Then, $\Omega_2-\Omega_2^\diamond>(VS_t+\varepsilon(1-\varepsilon)H_t\frac{\eta}{A}+c_t)(e_t^*-e^{\max})>(VB^{\max}+\varepsilon(1-\varepsilon)H_t\frac{\eta}{A}+c_t)(e_t^*-e^{\max})>0$.
        \item When $g_t^*<0$ and $g_t^\diamond>0$. Then, $\Omega_2-\Omega_2^\diamond>VS_tg_t^*-VB_tG_t^\diamond+(\varepsilon(1-\varepsilon)H_t\frac{\eta}{A}+c_t)(e_t^*-e^{\max})>(VB^{\max}+\varepsilon(1-\varepsilon)H_t\frac{\eta}{A}+c_t)(e_t^*-e^{\max})>0$.
      \end{itemize}
      Taking the above three cases into consideration, we have the conclusion that $e_t^*=e^{\max}$ when $VB^{\max}+c_t<-\varepsilon(1-\varepsilon)H_t\frac{\eta}{A}$.
\end{enumerate}
\end{IEEEproof}

\section{Proof of Theorem 1}
\begin{IEEEproof}
We will prove that the above inequalities are satisfied for all time slots by using mathematical induction method.
Since $T^{\min}\leq S_0\leq T^{\max}$, the above inequalities hold for $t$=0. Suppose the above-mentioned inequalities hold for the time slot $t$, we should verify that they still hold for the time slot $t$+1. The specific proof detail is given as follows.
\begin{itemize}
  \item If $\frac{VS^{\min}+b_t}{-\varepsilon(1-\varepsilon)\frac{\eta}{A}}-\Gamma<T_t\leq T^{\max}$. Then, the optimal HVAC decision is $e_t=0$ according to the Lemma 1. As a result, $T_{t+1}=\varepsilon T_{t}+(1-\varepsilon)T_t^{\text{out}}\leq \varepsilon T^{\max}+(1-\varepsilon)T^{\text{outmax}}\leq T^{\max}$, where (17) is incorporated. Similarly, $T_{t+1}\geq \frac{VS^{\min}+b_t}{-(1-\varepsilon)\frac{\eta}{A}}-\varepsilon \Gamma+(1-\varepsilon)T^{\text{outmin}}>T^{\min}$, where $\Gamma=\Gamma^{\max}$ is adopted.
  \item If $T^{\min}\leq T_t< \frac{VB^{\max}+c_t}{-\varepsilon(1-\varepsilon)\frac{\eta}{A}}-\Gamma$, then, the optimal HVAC decision is $e_t=e^{\max}$ according to the Lemma 1. Continually, $T_{t+1}\leq \frac{VB^{\max}+c_t}{-(1-\varepsilon)\frac{\eta}{A}}-\varepsilon \Gamma+(1-\varepsilon)(T^{\text{outmax}}+\frac{\eta}{A}e^{\max})<T^{\text{max}}$, where $\Gamma=\Gamma^{\min}$ is adopted.
      Similarly, $T_{t+1}\geq \varepsilon T^{\min}+(1-\varepsilon)(\frac{\eta}{A}e^{\max}+T^{\text{outmin}})\geq T^{\min}$, where (18) is incorporated.
  \item If $\frac{VB^{\max}+c_t}{-\varepsilon(1-\varepsilon)\frac{\eta}{A}}-\Gamma\leq T_t\leq \frac{VS^{\min}+b_t}{-\varepsilon(1-\varepsilon)\frac{\eta}{A}}-\Gamma$, $T_{t+1}\leq \frac{VS^{\min}+b_t}{-(1-\varepsilon)\frac{\eta}{A}}-\varepsilon\Gamma+(1-\varepsilon)(T^{\text{outmax}}+\frac{\eta}{A}e^{\max})\leq T^{\max}$, where $\Gamma=\Gamma^{\min}$ is adopted. Similarly, $T_{t+1}\geq \frac{VB^{\max}+c_t}{-(1-\varepsilon)\frac{\eta}{A}}-\varepsilon\Gamma+(1-\varepsilon)T^{\text{outmin}}\geq  T^{\min}$, where $\Gamma=\Gamma^{\max}$ is adopted.
\end{itemize}
\end{IEEEproof}

\section{Proof of Lemma 2}
\begin{IEEEproof}
Let ($e_t^*$,$x_t^*$,$y_t^*$,$g_t^*$) be the optimal solution of \textbf{P2}.
\begin{enumerate}
  \item When $Q_t+Z_t<VS^{\min}$, suppose $x_t^*>0$. To prove that the above assumption does not hold, we construct another solution ($e_t^*$,0,$y_t^*$,$g_t^\diamond$). According to power balance, we have $g_t^\diamond=g_t^*-x_t^*$. Let the optimal objective value of \textbf{P2} be $\Omega_2$ and the objective value with the new solution ($e_t^*$,0,$y_t^*$,$g_t^\diamond$) be $\Omega_2^\diamond$, respectively.
      Then, we can compare $\Omega_2$ with $\Omega_2^\diamond$ under different symbols of $g_t^\diamond$ and $g_t^*$ as follows.
      \begin{itemize}
        \item When $g_t^*<0$, we have $g_t^\diamond <0$ since $g_t^\diamond=g_t^*-x_t^*$ and $x_t^*>0$. Then, $\Omega_2-\Omega_2^\diamond=-(Q_t+Z_t)x_t^*+VS_tx_t^*=(VS_t-Q_t-Z_t)x_t^*>(VS^{\min}-Q_t-Z_t)x_t^*>0$.
        \item When $g_t^*>0$ and $g_t^\diamond>0$. Then, $\Omega_2-\Omega_2^\diamond=-(Q_t+Z_t)x_t^*+VB_tx_t^*=(VB_t-Q_t-Z_t)x_t^*>(VS^{\min}-Q_t-Z_t)x_t^*>0$.
        \item When $g_t^*>0$ and $g_t^\diamond<0$. Then, $\Omega_2-\Omega_2^\diamond>(VB_t-Q_t-Z_t)x_t^*>(VS_t-Q_t-Z_t)x_t^*>(VS^{\min}-Q_t-Z_t)x_t^*>0$.
      \end{itemize}
      Taking the above three cases into consideration, we have the conclusion that $x_t^*=0$ when $Q_t+Z_t<VS^{\min}$.
  \item When $Q_t+Z_t>VB^{\max}$, suppose $x_t^*<\min\{x^{\max},Q_t\}$. To prove that the above assumption does not hold, we construct another solution ($e_t^*$,$\min\{x^{\max},Q_t\}$,$y_t^*$,$g_t^\diamond$). According to power balance, we have $g_t^*-g_t^\diamond=x_t^*-\min\{x^{\max},Q_t\}$. Let the optimal objective value of \textbf{P2} be $\Omega_2$ and the objective value with the new solution ($e_t^*$,$\min\{x^{\max},Q_t\}$,$y_t^*$,$g_t^\diamond$) be $\Omega_2^\diamond$, respectively.
      Then, we can compare $\Omega_2$ with $\Omega_2^\diamond$ under different symbols of $g_t^\diamond$ and $g_t^*$ as follows.
      \begin{itemize}
        \item When $g_t^*>0$, we have $g_t^\diamond >0$ since $g_t^\diamond>g_t^*$. Then, $\Omega_2-\Omega_2^\diamond=(Q_t+Z_t-VB_t)(\min\{x^{\max},Q_t\}-x_t^*)>(Q_t+Z_t-VB^{\max})(\min\{x^{\max},Q_t\}-x_t^*)>0$.
        \item When $g_t^\diamond<0$ and $g_t^*<0$. Then, $\Omega_2-\Omega_2^\diamond=(Q_t+Z_t-VS_t)(\min\{x^{\max},Q_t\}-x_t^*)>(Q_t+Z_t-VB^{\max})(\min\{x^{\max},Q_t\}-x_t^*)>0$.
        \item When $g_t^*<0$ and $g_t^\diamond>0$. Then, $\Omega_2-\Omega_2^\diamond>VS_tg_t^*-VB_tg_t^\diamond+(Q_t+Z_t)(\min\{x^{\max},Q_t\}-x_t^*)>(Q_t+Z_t-VB^{\max})(\min\{x^{\max},Q_t\}-x_t^*)>0$.
      \end{itemize}
      Taking the above three cases into consideration, we have the conclusion that $x_t^*=\min\{x^{\max},Q_t\}$ when $Q_t+Z_t>VB^{\max}$.
\end{enumerate}
\end{IEEEproof}

\section{Proof of Theorem 2}
\begin{IEEEproof}
\begin{enumerate}
\item We prove the part 1 using the mathematical induction method. It can be observed that $Q_0<Q^{\max}$. Suppose we have $Q_t\leq Q^{\max}$, then, we need to prove that $Q_{t+1}\leq Q^{\max}$. If $Q_t\leq VB^{\max}$, $Q_{t+1}\leq Q_t+a_t\leq VB^{\max}+a^{\max}=Q^{\max}$. If $Q_t\geq VB^{\max}$, then, $x_t=\min\{x^{\max},Q_t\}$ according to the Lemma 2. Then, $Q_{t+1}\leq \max\{a^{\max},Q_t\}\leq Q^{\max}$. In summary, we have $Q_t\leq Q^{\max}$. As a result, \eqref{f_7} could be satisfied. Similarly, we can prove that $Z_t$ is bounded by $Z^{\max}=VB^{\max}+\xi$. The detail is omitted for brevity.
\item In section \ref{s3}-A, we know that maximum queueing delay is given by $D^{\max}=\lceil(Q^{\max}+Z^{\max})/\xi\rceil$. Taking the expressions of $Q^{\max}$ and $Z^{\max}$ into consideration, we have $D^{\max}=\left\lceil\frac{2VB^{\max}+a^{\max}+\xi}{\xi}\right\rceil$. Given the tolerant EV charging service delay $R$, we can obtain the minimum $\xi$. In summary, \eqref{f_8} could be satisfied under the proposed algorithm.
\end{enumerate}
\end{IEEEproof}

\section{Proof of Lemma 3}
\begin{IEEEproof}
Let ($e_t^*$,$x_t^*$,$y_t^*$,$g_t^*$) be the optimal solution of \textbf{P2},
\begin{enumerate}
  \item When $K_t>-VS^{\min}$, suppose $y_t^*>0$. To prove that the above assumption does not hold, we construct another solution ($e_t^*$,$x_t^*$,0,$g_t^\diamond$). According to power balance, we have $g_t^\diamond=g_t^*-y_t^*$. Let the optimal objective value of \textbf{P2} be $\Omega_2$ and the objective value with the new solution ($e_t^*$,$x_t^*$,0,$g_t^\diamond$) be $\Omega_2^\diamond$, respectively.
      Then, we can compare $\Omega_2$ with $\Omega_2^\diamond$ under different symbols of $g_t^\diamond$ and $g_t^*$ as follows.
      \begin{itemize}
        \item When $g_t^*<0$, we have $g_t^\diamond <0$ since $g_t^\diamond=g_t^*-y_t^*$ and $y_t^*>0$. Then, $\Omega_2-\Omega_2^\diamond=(K_t+VS_t)y_t^*\geq (K_t+VS^{\min})y_t^*>0$.
        \item When $g_t^*>0$ and $g_t^\diamond>0$. Then, $\Omega_2-\Omega_2^\diamond=(K_t+VB_t)y_t^*>(K_t+VS_t)y_t^*>(K_t+VS^{\min})y_t^*>0$.
        \item When $g_t^*>0$ and $g_t^\diamond<0$. Then, $\Omega_2-\Omega_2^\diamond>K_t y_t^*+VB_tg_t^*-VS_t g_t^\diamond>(K_t+VS_t)y_t^*>(K_t+VS^{\min})y_t^*>0$.
      \end{itemize}
      Taking the above three cases into consideration, we have the conclusion that $y_t^*\leq 0$ when $K_t>-VS^{\min}$.

  \item When $K_t<-VB^{\max}$, suppose $y_t^*<0$. To prove that the above assumption does not hold, we construct another solution ($e_t^*$,$x_t^*$,0,$g_t^\diamond$). According to power balance, we have $g_t^\diamond=g_t^*-y_t^*$. Let the optimal objective value of \textbf{P2} be $\Omega_2$ and the objective value with the new solution ($e_t^*$,$x_t^*$,0,$g_t^\diamond$) be $\Omega_2^\diamond$, respectively.
      Then, we can compare $\Omega_2$ with $\Omega_2^\diamond$ under different symbols of $g_t^\diamond$ and $g_t^*$ as follows.
      \begin{itemize}
        \item When $g_t^*>0$, we have $g_t^\diamond >0$ since $g_t^\diamond=g_t^*-y_t^*$ and $y_t^*<0$. Then, $\Omega_2-\Omega_2^\diamond=(K_t+VB_t)y_t^*\geq (K_t+VB^{\max})y_t^*>0$.
        \item When $g_t^\diamond<0$ and $g_t^*<0$. Then, $\Omega_2-\Omega_2^\diamond=(K_t+VS_t)y_t^*>(K_t+VB^{\max})y_t^*>0$.
        \item When $g_t^\diamond>0$ and $g_t^*<0$. Then, $\Omega_2-\Omega_2^\diamond>(K_t+VB_t)y_t^*+V(B_t-S_t) g_t^\diamond>(K_t+VB_t)y_t^*>(K_t+VB^{\max})y_t^*>0$.
      \end{itemize}
      Taking the above three cases into consideration, we have the conclusion that $y_t^*\geq 0$ when $K_t<-VB^{\max}$.
\end{enumerate}
\end{IEEEproof}

\section{Proof of Theorem 3}
\begin{IEEEproof}
We will prove that the above inequalities are satisfied for all time slots by using mathematical induction method.
Since $G^{\min}\leq G_0\leq G^{\max}$, the above inequalities hold for $t$=0. Suppose the above-mentioned inequalities hold for the time slot $t$, we should verify that they still hold for the time slot $t$+1. The specific proof detail is given as follows.
\begin{itemize}
  \item If $-VS^{\min}-\alpha<G_t\leq G^{\max}$. Then, the optimal ESS decision is $y_t^*\leq 0$ according to the Lemma 3. As a result, $G_{t+1}=G_t+y_t^*\leq G^{\max}$. Similarly, $G_{t+1}\geq -VS^{\min}-\alpha-u^{\text{dmax}}>G^{\min}$, where $\alpha=\alpha^{\max}$ is adopted.
  \item If $G^{\min}\leq G_t< -VB^{\max}-\alpha$, then, the optimal ESS decision is $y_t^*\geq 0$ according to the Lemma 1. Continually, $G_{t+1}\leq -VB^{\max}-\alpha+u^{\text{cmax}}\leq G^{\max}$, where $\alpha=\alpha^{\min}$ is adopted.
      Similarly, $G_{t+1}\geq G^{\min}+y_t^* \geq G^{\min}$.
  \item If $-VB^{\max}-\alpha \leq G_t\leq -VS^{\min}-\alpha$, $G_{t+1}\leq -VS^{\min}-\alpha+u^{\text{cmax}} \leq G^{\max}$, where $\alpha=\alpha^{\min}$ is adopted. Similarly, $G_{t+1}\geq -VB^{\max}-\alpha-u^{\text{dmax}}\geq G^{\min}$, where $\alpha=\alpha^{\max}$ is adopted.
\end{itemize}
\end{IEEEproof}

\section{Proof of Theorem 4}
\begin{IEEEproof}
To prove the performance of the proposed algorithm, we first define some equations as follows, i.e., $\overline{a}=\mathop {\lim \sup }\limits_{N \to \infty } \frac{1}{N-1}\sum\nolimits_{t = 0}^{N - 2} \mathbb{E}\{a_t\}$, $\overline{x}=\mathop {\lim \sup }\limits_{N \to \infty } \frac{1}{N-1}\sum\nolimits_{t = 0}^{N - 2} \mathbb{E}\{x_t\}$, $\overline{y}=\mathop {\lim \sup }\limits_{N \to \infty } \frac{1}{N-1}\sum\nolimits_{t = 0}^{N - 2} \mathbb{E}\{y_t\}$, $\overline{e}=\mathop {\lim \sup }\limits_{N \to \infty } \frac{1}{N-1}\sum\nolimits_{t = 0}^{N - 2} \mathbb{E}\{e_t\}$. Then, we have $\overline{y}=0$ based on the constraint \eqref{f_10}. Similarly, we have $\frac{A}{\eta}(T^{\min}-T^{\text{outmax}})\leq \overline{e}\leq \frac{A}{\eta}(T^{\max}-T^{\text{outmin}})$ based on the constraint \eqref{f_1}. In addition, based on the constraint \eqref{f_4}, we have $\overline{a}\leq \overline{x}$. Then, we consider the following optimization problem as follows,
\begin{subequations}\label{f_th4}
\begin{align}
(\textbf{P3})~&\min~\mathop {\lim\sup}\limits_{N \to \infty}\frac{1}{N-1}\sum\limits_{t=0}^{N-2} \mathbb{E}\{\Phi_{1,t}+\Phi_{2,t}\}  \\
s.t.&~(3),(6),(9),(10),(13),\\
&\frac{A}{\eta}(T^{\min}-T^{\text{outmax}})\leq \overline{e}\leq \frac{A}{\eta}(T^{\max}-T^{\text{outmin}}),\\
&\overline{a}\leq\overline{x},\\
&\overline{y}=0;
\end{align}
\end{subequations}
Note that (2),(4) are replaced by (39c), (5),(8) are replaced by (39d),(11),(12) are replaced by (39e). Since any feasible solution of \textbf{P1} is also feasible to \textbf{P3}, we have $y_2\leq y_1$, where $y_2$ and $y_1$ are the optimal objective values of \textbf{P3} and \textbf{P1}, respectively. Using the Theorem 4.5 in \cite{Neely2010}, the conclusion could be obtained similarly, i.e., if purchasing/selling electricity prices $B_t$/$S_t$, outdoor temperatures $T_t^{\text{out}}$, renewable generation outputs $r_t$, EV electrical demand $a_t$, the most comfortable temperature level $T_{t+1}^{\text{ref}}$, and home occupancy state $T_{t+1}^{\text{ref}}$ are i.i.d. over slots and \textbf{P3} is feasible, there exists a stationary, randomized policy that takes control decision ($x_t^*,e_t^*,y_t^*,g_t^*$) purely as a function of current system observation parameters and provides the following performance guarantee, i.e., $\mathbb{E}\{\Phi_{1,t}^*+\Phi_{2,t}^*\}\leq y_2$, $\mathbb{E}\{y_t^*\}=0$, $\mathbb{E}\{a_t^*\}\leq \mathbb{E}\{x_t^*\}$, and $\frac{\eta }{A}{\mathbb{E}\{ e_t^*\}}\leq T^{\max}-T^{\text{outmin}}$. Continually, when using the proposed algorithm, we have
\begin{align} \label{f_th41}
&\Delta Y_t=\Delta_t + V\mathbb{E}\{ {\Phi_{1,t}+\Phi_{2,t}|\boldsymbol{\Psi_t}} \}\nonumber\\
\leq& \sum\nolimits_{l=1}^4\Omega_l+\mathbb{E}\{K_t y_t^*-(Q_t+Z_t)x_t^*|\boldsymbol{\Psi_t}\} \\
&+\mathbb{E}\{\varepsilon (1 - \varepsilon ){H_t}(\Gamma  + T_t^{\text{out}} + \frac{\eta }{A}{e_t^*})|\boldsymbol{\Psi_t}\}\nonumber \\
&+V\mathbb{E}\{ {\Phi_{1,t}^*+\Phi_{2,t}^*|\boldsymbol{\Psi_t}} \}, \nonumber \\
\leq &\sum\nolimits_{l=1}^{4}\Omega_l+Vy_2+\Upsilon,  \\
\leq &\Theta+Vy_1,
\end{align}
where $\Upsilon=\varepsilon (1-\varepsilon )(T^{\max}+\Gamma)(T^{\max}+\Gamma+(T^{\text{outmax}}-T^{\text{outmin}}))$, $\Theta=\sum\nolimits_{l=1}^{4}\Omega_l+\Upsilon$, (40) holds due to that the proposed algorithm minimizes the upper bound given in the right-hand-side of the \emph{drift-plus-penalty} term over all other control strategies, including the optimal stationary and randomized control strategy; (41) is obtained by incorporating the results of a stationary, randomized control strategy associated with \textbf{P3}. In addition, $H_t\leq T^{\max}+\Gamma$, $T_t^{\text{out}}\leq T^{\text{outmax}}$. By arranging the both sides of the above equations, we have $\mathbb{E}\{\Delta_t\} + V\mathbb{E}\{\Phi_{1,t}+\Phi_{2,t}\} \leq \Theta+Vy_1$. Continually, we have $V {\sum\nolimits_{t = 0}^{N - 2} {\mathbb{E}\{\Phi_{1,t}+\Phi_{2,t}\}}} \leq \Theta (N-1)+V(N-1)y_1 - \mathbb{E}\{L_{N-1}\}+\mathbb{E}\{L_0\}$. Dividing both side by $V(N-1)$, and taking a \text{lim sup} of both sides. Then, let $N \to \infty$, we have $\mathop {\lim \sup }\nolimits_{N \to \infty } \frac{1}{N-1}{\sum\nolimits_{t = 0}^{N-2} {\mathbb{E}\{\Phi_{1,t}+\Phi_{2,t}\}}} \le y_1+\frac{\Theta}{V}$, which completes the proof.
\end{IEEEproof}

\end{document}